\documentclass[aps,prb,twocolumn,groupedaddress]{revtex4-2} 

\usepackage[]{graphicx}
\usepackage[dvipsnames]{xcolor}
\usepackage[colorlinks=true,citecolor=NavyBlue,linkcolor=RubineRed,urlcolor=Cerulean]{hyperref}
\usepackage[]{amsmath}
\usepackage[]{amssymb}

\def\bra#1{\mathinner{\langle{#1}|}}
\def\ket#1{\mathinner{|{#1}\rangle}}
\def\braket#1{\mathinner{\langle{#1}\rangle}}

\newcommand\idx[3]{%
  {\vphantom{#2}}#1\hspace{-0.1em}{#2}#3%
}
\def\braketo#1#2{\idx{^{#2}}{\braket{#1}}{^{#2}}}

\def\abs#1{\mathinner{|{#1}|}}

\def\proj#1{\ket{#1}\bra{#1}}
\def\op#1#2{\ket{#1}\bra{#2}}

\def\comm#1#2{\mathinner{[{#1},{#2}]}}

\DeclareMathOperator{\I}{\openone}

\begin{document}

\title{Spin Structure and Resonant Driving of Spin-1/2 Defects in SiC}

\author{Benedikt Tissot}
\email[]{benedikt.tissot@uni-konstanz.de}

\author{Guido Burkard}
\email[]{guido.burkard@uni-konstanz.de}

\affiliation{Department of Physics, University of Konstanz, D-78457 Konstanz, Germany}


\begin{abstract}
Transition metal (TM) defects in silicon carbide have favorable spin coherence properties and are suitable as quantum memory for quantum communication.
  To characterize TM defects as quantum spin-photon interfaces, we model defects that have one active electron with spin 1/2 in the atomic $D$ shell.
  The spin structure, as well as the magnetic and optical  resonance properties of the active electron emerge from the interplay of the crystal potential and spin-orbit coupling and are described by a general model derived using group theory.
  We find that the spin-orbit coupling leads to additional allowed transitions and a modification of the $g$-tensor.
To describe the  dependence of the Rabi frequency on the magnitude and direction of the static and driving fields,  we derive an effective Hamiltonian.
  This theoretical description can also be instrumental to perform and optimize spin control in TM defects.

\end{abstract}

\maketitle
\section{Introduction}

To implement quantum communication it is necessary to transfer quantum information between stationary and mobile carriers~\cite{gisin07}.
Photons are by far the most adequate choice as the mobile carriers of quantum information.
Fiber-optic cables spread over the globe make use of the efficient long range transmission of light.
The frequencies within the electromagnetic spectrum that are transmitted  efficiently by optical fibers lie within the telecommunication bands.
There is a substantial interest to develop quantum systems that can emit photons within these frequency bands to harness the available infrastructure for quantum communication~\cite{kimble08}.
Transition metal (TM) defects in silicon carbide (SiC) are promising candidates as a platform in an industrially established material for emitters in such a frequency range~\cite{spindlberger19,wolfowicz20,gilardoni20}.
Particularly encouraging are the observed long spin relaxation times $T_1$ for molybdenum defects 
exceeding seconds~\cite{gilardoni20}, and inhomogeneous dephasing times around $T_2^*\approx 0.3\,\mu{\rm s}$ \cite{bosma18}.

\begin{figure}
\includegraphics[width=\columnwidth]{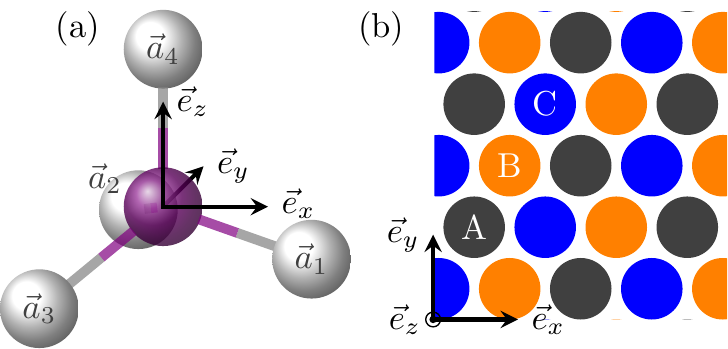}%
\caption{\label{fig:sym} Lattice structure of SiC. (a) Schematic of one lattice site occupied by Si (purple) with the bonding C atoms (white).
  These nearest neighbors fulfil the symmetry $T_d$ of the regular tetrahedron.
  Typically, the substitutional TM atom occupies a Si site such as the one shown in purple.
  (b) The three possible layers ($A$, $B$, $C$) are shown viewed from the top where balls of one color denote lattice sites in the same layer. 
  The SiC lattice can be built up by stacking tetrahedrally bonded Si-C bilayers, where
  each layer can only be followed by  one of the other two types.
  (Quasi-) hexagonal sites lie in a layer surrounded by layers of the same type (e.g., $B$ in $ABA$) while (quasi-) cubic sites are surrounded by different layers (e.g., $B$ in $ABC$).
  SiC polytypes other than cubic $3C$-SiC break the $T_d$ symmetry.}
\end{figure}

The energy level structure of vanadium (V) and molybdenum (Mo) defects in SiC as well as their ground state spin properties without sub-level structure are already well understood~\cite{kunzer93,reinke93,kaufmann97,bosma18,csore19,spindlberger19,wolfowicz20,gilardoni20}.
However, in order to fully understand the selection rules, allowed and forbidden transitions, and Rabi frequencies for arbitrary orientations of the static and oscillatory (electric or magnetic) fields, more details of the level structure are required.
Inspired by studies  of the nitrogen vacancy centre in diamond \cite{awschalom18} employing group theory \cite{lenef96,tamarat08,doherty11,maze11,doherty13} that shares the symmetry with the TM defects in SiC studied in this article,
we also employ group-theoretical methods to derive a Hamiltonian for an active electron localized in a $d$ orbital of the TM defect (V or Mo) that possesses $C_{3v}$ symmetry imposed by the crystal field (Fig.~\ref{fig:sym}).
We analytically compute the defect energy levels as given by the eigenvalues of this Hamiltonian in the absence of external fields.

In the presence of a non-zero static external (magnetic or electric) field the symmetry of the system can be reduced and therefore previously used naive group theoretical arguments, e.g., selection rules based on the symmetry, are no longer applicable.
To overcome this obstacle we use a Schrieffer-Wolff transformation to derive an effective Hamiltonian
which is compatible with ground state Zeeman Hamiltonians derived previously~\cite{kaufmann97} but
has the benefit that it directly links the effective $g$-factors to the spin-orbit coupling.
This procedure also shows that some matrix elements inside an orbital doublet that would not vanish in the most general case, do vanish because the states originate from an atomic $d$ orbital.

For static magnetic fields along the high symmetry axis of the crystal we derive selection rules within first and second order perturbation theory in the spin-orbit interaction, making it possible to relate the magnitude of various transitions based on a small set of system properties.
Furthermore, the effective Hamiltonian describes how a static magnetic field breaking the $C_{3v}$ point symmetry mixes states of different irreducible representations (irreps) and thus changes the selection rules.
To understand magnetic and optical resonance in TM defects in more detail we study the dependence of the Rabi frequency on the magnetic field direction and provide examples where otherwise forbidden transitions are allowed for static magnetic fields that break the point symmetry.

The remainder of this paper is structured as follows:
In Section~\ref{sec:model} we introduce the Hamiltonian describing the system.  
In Section~\ref{sec:results} we treat the model where
we first show that the reduced problem can be solved analytically in absence of an external field
and then (Section~\ref{sec:SW}) use the Schrieffer-Wolff transformation to study the influence of external fields.
In Section~\ref{sec:resonance} we combine these insights and present the selection rules as well as the Rabi frequencies for the various spin transitions in the TM defect.
We conclude our work in Section~\ref{sec:conclusion}.

\section{Model\label{sec:model}}
We consider spin-$1/2$ defects in SiC where neutral V$^0$ or positively charged  Mo$^+$ substitute a silicon atom~\cite{csore19}, resulting in one active electron in a $d^1$ atomic orbital.
These defects are sometimes referred to as V$^{4+}$ and Mo$^{5+}$ in the literature~\cite{kaufmann97,bosma18,wolfowicz20}.
Our theory is based on the $d^1$ character of the electronic state and the $C_{3v}$ point symmetry of the surrounding crystal.
Since TM defects in SiC are not the only systems that fulfill these properties, the theory is also applicable to other systems, e.g., copper impurities in ZnO~\cite{dietz63}.

The full model Hamiltonian we consider takes the form~\cite{dresselhaus10}
\begin{align}
  \label{eq:Htotal}
  H = H_{\mathrm{TM}} + V_{\mathrm{cr}} + {H}_{\mathrm{so}} + {H}_{\mathrm{hf}} + {H}_{z} + V_{\mathrm{el}} .
\end{align}
The dominating part is the atomic Hamiltonian $H_{\mathrm{TM}} = p^2/(2 m_e) + V_{\mathrm{TM}}$ where $\vec{p}$ denotes the momentum, $m_e$ the mass of the active electron,
and the coupling of the Coulomb potential of the TM atom to the electron charge $-e$ given by $V_{\mathrm{TM}}$.
This potential only depends on the distance $r = |\vec{r}|$ between the electron and the defect placed at the origin, localizes the active electron at the defect site, and separates the energy of the atomic $D$ shell from the remaining spectrum.
The Coulomb potentials of the crystal atoms
couple to the charge of the active electron, resulting in the crystal potential $V_{\mathrm{cr}}$ that breaks the spherical symmetry of the defect atom and reduces it to the $C_{3v}$ point group.

An electron in a $d$ orbital state has non-zero angular momentum $l=2$
inducing a magnetic field in the electron rest frame.
This relativistic effect is taken into account in the spin-orbit coupling Hamiltonian~\cite{thomas26,dresselhaus10}
\begin{align}
  \label{eq:HsoGeneral}
  H_{\mathrm{so}} = \frac{\hbar}{2 m_e^2 c^2} \left\{ \nabla \left[ V_{\mathrm{TM}} + V_{\mathrm{cr}} \right] \times \vec{p} \right\} \cdot \vec{S}, 
\end{align}
with the electron spin vector operator $\vec{S} = \vec{\sigma}/2$ in units of the reduced Planck constant $\hbar$ given by half the Pauli vector $\vec{\sigma}$,
and the speed of light in vacuum $c$.
Considering that the gradient of the Coulomb potentials
$\nabla \left[ V_{\mathrm{TM}} + V_{\mathrm{cr}} \right]$
transforms like the vector $\vec{r}$
the complete orbital part transforms like the orbital angular momentum operator $\vec{L} = \vec{r} \times \vec{p} / \hbar$ (in units of $\hbar$)~\cite{dresselhaus10}.
For the free ion, where $V_{\mathrm{cr}} = 0$, the intact spherical symmetry  leads to
  $H_{\mathrm{so}} = \lambda_0 \vec{L} \cdot \vec{S}$,
with the free ion coupling constant ${\lambda_0={\mu_0 Z \mu_B^2}/{(2 \pi r^3)}}$ 
expressed via the vacuum permeability $\mu_0$,
the atomic number $Z$, and the Bohr magneton $\mu_B$
and depending on the electronic configuration via $1/r^3$.

The hyperfine Hamiltonian $H_{\mathrm{hf}}$ models the interaction of the active electron's spin with nearby nuclear spins.
For an electron localized in an atomic $d$ orbital at the defect site, we expect the interaction with the nuclear spin of the defect site to be the dominating contribution.
The hyperfine coupling strength is~\cite{coish09}
$a = g_s \mu_{B} \mu_0 g_N \mu_N/4\pi r^{3}$,
with the nuclear magneton $\mu_N$ and $g$-factor $g_N$ (of the defect) as well as the electron $g$-factor $g_s$.
The most common V isotope is $^{51}$V with an abundance larger than $99\%$ and nuclear spin ${7}/{2}$,
while Mo only has approximately $25\%$ combined natural abundance for isotopes $^{95}$Mo and $^{97}$Mo with nuclear spin ${5}/{2}$ and the remaining naturally occurring isotopes have nuclear spin zero~\cite{audi03,meija16}.
We compare the orders of magnitude by the free ion values
$\abs{{a}}/{\lambda_{0}} = {g_s \abs{g_N \mu_N}}/{2 \mu_B Z} \approx 5.5 \times 10^{-6}$,
where the approximate value is given for experimental values of $g_N \mu_N$ for V~\cite{wolfowicz20}.
Compared to V, Mo has the larger atomic number resulting in a larger spin-orbit coupling constant~\cite{koseki19}.
Furthermore, the isotopes with non-zero nuclear spin of Mo have a smaller $\abs{\mu_N g_N}$ compared to the relevant isotope of V~\cite{stone05}.
Therefore, the ratio of the hyperfine and the spin-orbit coupling strength for V gives the upper bound for the two TM atoms.

The Zeeman Hamiltonian describes the coupling of the electron spin $\vec{S}$, electron angular momentum $\vec{L}$ and nuclear spin $\vec{I}$ to a uniform external magnetic field $\vec{B}$.
This term is given by
\begin{align}
  \label{eq:HzGeneral}
  H_{z} = \mu_B  \left( g_s \vec{B} \cdot \vec{S} + \vec{B} \cdot \vec{L} \right) - \mu_N g_N \vec{B} \cdot \vec{I} .
\end{align}
Again using the experimental values of $g_N \mu_N$ for V~\cite{wolfowicz20} we find ${g_{N} \mu_N}/{\mu_B} \approx 10^{-4}$.
From that we conclude $H_{z}$ is dominated by the coupling of the magnetic field to the electron spin and angular momentum.

The active electron can also couple to external electric fields, resulting in the potential $V_{\mathrm{el}}$.
Assuming that the electric field is uniform over the scale of the defect the term is given by
\begin{align}
  \label{eq:VelGeneral}
  V_{\mathrm{el}} = e \vec{E} \cdot \vec{r},
\end{align}
where $\vec{E}$ is the external electric field.

Using these considerations we order the magnitudes of the different contributions
\begin{align}
  \label{eq:termmagnitudes}
H_{\mathrm{TM}} \gg V_{\mathrm{cr}} \gg H_{\mathrm{so}} \gg H_{\mathrm{hf}}.
\end{align}
In the following we will concentrate on the case where we can neglect the hyperfine interaction,
given by static magnetic fields $\abs{\vec{B}} \gg a/\mu_B \approx 3 \, $mT.
For electric fields well below the breakdown field strength of SiC $\abs{\vec{E}} \ll 3\,$MV/cm~\cite{yamaguchi18} we estimate $H_{\mathrm{so}} \gg V_{\mathrm{el}}$ and for magnetic fields smaller than
$B_{\mathrm{so}} = \lambda_0 / \mu_B \approx 531 \, \mathrm{T}$
we have $H_{\mathrm{so}} \gg H_{z}$.
The approximate values are given for the free ion value of the spin-orbit coupling constant $\lambda_0$ for V~\cite{kaufmann97} and therefore give the lower bound for the two TMs.

\subsection{Symmetry}

SiC consists of two constituents, silicon (Si) and carbon (C).
Each atom of the lattice is tetragonally bound to four atoms of the other constituent,
giving rise to the tetragonal point symmetry $T_d$ of one crystal site with its nearest neighbors, see Fig.~\ref{fig:sym}(a).

To take the remaining atoms into account, it is necessary to consider the crystal structure of
SiC which can be described in terms of polytypeism~\cite{morkoc94}.
In the following, we describe the crystal structure by the stacking order of tetrahedrally bonded Si-C bilayers and refer to the stacking axis as the crystal axis.
Due to the tetrahedral bonding, there are only three inequivalent types of layers, 
distinguished by the  positions of the Si atoms (lattice sites) in the stacking plane [Fig.~\ref{fig:sym}(b)].
In particular, layers of the same type cannot be directly on top of each other due to the bonding of the atoms in the crystal structure,
leaving only two possibilities to stack three layers.

Labeling the inequivalent layers $A$,$B$ and $C$,
the first possible stacking has the form $ABC$ where sites in the $B$ layer have a zinkblende bonding along the crystal axis and are called (quasi-) cubic sites.
The zinkblende bonding is compatible with the $T_d$ symmetry of the nearest neighbors.
The other possible stacking has the form $ABA$ where the bonding along the crystal axis is given by wurtzite bonding and sites in the $B$ layer are called (quasi-) hexagonal.
Hexagonal layers reduce the symmetry of point defects from $T_d$ to $C_{3v}$.

The various polytypes of SiC are combinations of hexagonal and cubic layers.
All polytypes apart from cubic $3C$-SiC have at least one hexagonal layer and therefore defect sites in these crystals fulfill $C_{3v}$ symmetry,
where the high symmetry axis coincides with the crystal axis.
In addition to these symmetries, in the absence of a magnetic field, the system is also invariant under time inversion.

\begin{figure}
\includegraphics[width=\columnwidth]{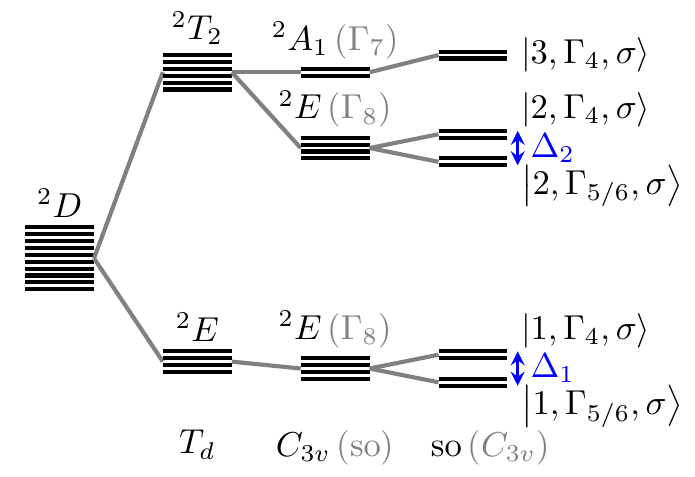}
\caption{\label{fig:Ediag}The energy levels of a $d^1$ defect in SiC.
  The spherical symmetry of the defect is reduced to the $T_d$ point symmetry by its nearest neighbors.
  This symmetry is further reduced by atoms farther away to $C_{3v}$ in SiC polytypes apart from $3C$-SiC.
  The spin-orbit coupling splits the levels further, producing the fine structure given by 5 Kramers doublets.
  We show the spin-orbit splittings of the orbital doublets in blue, which to second order can be found in Eq.~\eqref{eq:Deltai}.
  If one considers the case where the spin-orbit splitting is larger than the splitting due to the symmetry reduction to $C_{3v}$ we obtain the same diagram but with $A_1 \to \Gamma_7$, $E \to \Gamma_8$ which are irreps of the $T_d$ double group of dimensions 2 and 4, respectively.}
\end{figure}
We now summarize the energy level structure that follows directly from group theory for a $d^1$-orbital state when the spherical symmetry is reduced to $T_d$ and then to $C_{3v}$, including spin-orbit coupling~\cite{dietz63,kunzer93,reinke93,kaufmann97,bosma18,spindlberger19,wolfowicz20,csore19,gilardoni20}, see Fig.~\ref{fig:Ediag}.
A $d$ orbital corresponds to the $\Gamma_{l=2}$ irrep of the full rotation group.
In the $T_d$ group the $\Gamma_{l=2}$ representation is composed of the three-dimensional irrep $T_1$ and the two-dimensional irrep $E$, hence the $d$ orbital is split into a doublet and a triplet in a $T_d$ symmetric potential.
While the doublet does not split further in $C_{3v}$, the triplet $T_1$ splits into another doublet $E$ and a singlet corresponding to the irrep $A_1$.

The spin-orbit coupling  renders
the orbital singlet of $C_{3v}$ into a Kramers doublet (KD),
a degenerate pair of states which are connected by time inversion,
transforming according to the spinor representation $\Gamma_4$.
The doublets of $C_{3v}$ split into two KDs corresponding to the irrep $\Gamma_4$ and the combined irreps  $\Gamma_{5/6} = \Gamma_5 \oplus \Gamma_6$, respectively.
The same final structure arises when first considering the spin-orbit coupling and then the symmetry reduction to $C_{3v}$.
Time-reversal symmetry protects the states of the same KD from being coupled to each other by operators that also fulfill time-reversal symmetry~\cite{dresselhaus10}, such as the position $\vec{r}$ and thus $V_{\mathrm{el}}$, as well as even orders of $\vec{L}$, $\vec{S}$ and $\vec{p}$ (combined).

\subsection{\texorpdfstring{$C_{3v}$}{C3v} Symmetric Hamiltonian}
To derive a general Hamiltonian for one active electron in the $d$ orbital subspace we use the basis
\begin{align}
  \label{eq:base}
  \ket{m} \ket{\sigma} = \ket{l=2,m} \ket{\sigma},
\end{align}
with the spherical harmonics $\ket{l=2,m}$ ($m =  -2,-1,0,1,2 $)
and the spinor $\ket{\sigma}$ ($\sigma =  \: \uparrow, \downarrow $)
for the $z$-axis parallel to the crystal axis.
The full electronic wavefunction $\ket{\Psi_m}$ in addition contains the radial part according to the main quantum number $n=3$ (V) or $n=4$ (Mo).
Due to effects such as covalency and the Jahn-Teller effect, the  states $\ket{\Psi_m}$   can have contributions from other orbital states~\cite{ham65,ham68,csore19}.
These effects are taken into account by a proportionality constant which describes the ratio between 
the \emph{effective} matrix element $\braket{m|O|m'}$ and full matrix element $\braket{\Psi_m|O|\Psi_{m'}}$.
A framework to derive operators that can take this into account while having the appropriate transformation properties is given by the Wigner-Eckart theorem where the proportionality constants correspond to so called reduced matrix elements~\cite{cornwell97}.

The transformation properties of the basis states and operators as well as the implications of these for the Wigner-Eckart theorem are given in Appendices~\ref{sec:symgroups}~to~\ref{sec:HoSphHarm}.
The resulting Hamiltonian is described in the following and generalizes previously used Hamiltonians derived by ligand or crystal field theory~\cite{dietz63,kaufmann97}.

\subsubsection{Crystal Eigenstates}
The purely orbital Hamiltonian 
$H_{o} = H_{\mathrm{TM}} + V_{\mathrm{cr}}$
which transforms according to the irrep $A_1$ of $C_{3v}$ and is time-reversal symmetric,
can be written as 
\begin{align}
  \label{eq:Hcr}
  H_{o} = \epsilon_3 \proj{0} + \sum_{i=1,2} \epsilon_i \left( \proj{+_i} + \proj{-_i} \right)
\end{align}
inside the subspace of the $d$ orbitals.
The orbital doublet states are given by
\begin{equation}
  \label{eq:HcrStates}
  \begin{aligned}
  \ket{\pm_1} = & \cos(\phi) \ket{\pm 1} \mp \sin(\phi) \ket{\mp 2} , \\
  \ket{\pm_2} = & - \sin(\phi) \ket{\pm 1} \mp \cos(\phi) \ket{\mp 2},
  \end{aligned}
\end{equation}
where the mixing angle $\phi$ describes the admixture of states that transform equally under $C_{3v}$.
The states $\ket{\pm_i}$ transform the same as $\ket{\pm 1}$ under the symmetry operations of $C_{3v}$ and time inversion ${T}_{\mathrm{inv}} \ket{\pm_i} = - \ket{\mp_i}$.

\subsubsection{Spin-Orbit Coupling and Zeeman Term}
The spin-orbit (Zeeman) Hamiltonian transforms like the scalar product of the angular momentum operator with the spin operator $\vec{L} \cdot \vec{S}$ (magnetic field $\vec{L} \cdot \vec{B}$).
This implies that the spin-orbit Hamiltonian is time-reversal symmetric, while the Zeeman Hamiltonian is not.
Using projection operators on the eigenspaces of $H_{\mathrm{cr}}$,
\begin{align}
  \label{eq:proj}
  & P_i = \proj{+_i} + \proj{-_i} ,
  & P_3 = \proj{0},
\end{align}
for $i = 1,2$, we can write the spin-orbit Hamiltonian as
\begin{align}
  \label{eq:Hso}
  P_i H_{\mathrm{so}} P_j = & \vec{S} \cdot \tilde{\Lambda}_{ij} \cdot P_i \vec{L} P_j,
\end{align}
where now $i=1,2,3$ and
$\tilde{\Lambda}_{ij}=\tilde{\Lambda}_{ji}= \operatorname{diag}(\tilde{\lambda}_{\perp,ij},\tilde{\lambda}_{\perp,ij},\tilde{\lambda}_{\parallel,ij})$.
Analogously, the Zeeman term is
\begin{align}
  \label{eq:Hz}
  P_i H_{z} P_j = & \delta_{ij} \mu_{B} g_{s} \vec{B} \cdot \vec{S} +
                    \mu_{B} \vec{B} \cdot \tilde{R}_{ij} \cdot P_i \vec{L} P_j,
\end{align}
with $\tilde{R}_{ij} = \tilde{R}_{ji} = \operatorname{diag}(\tilde{r}_{\perp,ij},\tilde{r}_{\perp,ij},\tilde{r}_{\parallel,ij})$.
The tensors $\tilde{R}$ and $\tilde{\Lambda}$ take the anisotropy of the reduction factors for the spin-orbit coupling $\tilde{\lambda}_{k,ij}/\lambda_0$ and the orbital Zeeman term $\tilde{r}_{k,ij}$ into account.

In the employed basis [Eq.~\eqref{eq:HcrStates}] the non-zero matrix elements of the orbital angular momentum operators $L_k$ are given by
$l_{\parallel,ij} = \pm \braket{\pm_i|L_z|\pm_j}$,
$l_{\perp,12} = \pm \braket{\mp_2 | L_x |\pm_1} = \mathrm{i} \braket{\mp_2 | L_y |\pm_1}$,
$l_{\perp,i3} = \braket{\pm_i|L_x|0} = \pm \mathrm{i} \braket{\pm_i|L_y|0}$,
with $i,j = 1,2 $.
Since $\vec{L}$ is Hermitian $l_{k,ij} = l_{k,ji}$ and the unique matrix elements in terms of the orbital mixing angle $\phi$ are given in Table~\ref{tab:sfacts}.
\begin{table}
  \caption{\label{tab:sfacts} The $l$-factors introduced in the main text, corresponding to the non-zero matrix elements of the angular momentum operators $L_k$ in the crystal eigenbasis.}
  \begin{ruledtabular}
    \begin{tabular}{l | rrrr}
      $i$ & $l_{\perp,i2}$ & $l_{\perp,i3}$ & $l_{\parallel,i1}$ & $l_{\parallel,i2}$  \\ \hline
      $1$ & $1$ & $\frac{\sqrt{6}}{2} \cos(\phi)$  & $ \frac{1}{2} \left[ 3 \cos(2\phi) - 1 \right]$ & $- \frac{3}{2} \sin(2\phi)$ \\
      $2$ & 0 & $-\frac{\sqrt{6}}{2}\sin(\phi)$ & 0 & $- \frac{1}{2} \left[ 3 \cos(2\phi) + 1 \right]$ \\
    \end{tabular}
  \end{ruledtabular}
\end{table}
From now on we will use the combined parameters $\lambda_{k,ij} = l_{k,ij} \tilde{\lambda}_{k,ij}$ and $r_{k,ij} = l_{k,ij} \tilde{r}_{k,ij}$.

\subsubsection{External Electric Field}
The coupling to external electric fields has the form shown in Eq.~\eqref{eq:VelGeneral} and is time-reversal symmetric for a static field.
Due to the shared symmetry of $x,y$ and $-L_{y},L_{x}$ we can use the orbital angular momentum operators to express
\begin{equation}
    \label{eq:Vel}
    \begin{aligned}
  P_i V_{\mathrm{el}} P_j = & (-1)^{\Theta(i-j)} \mathrm{i} \tilde{\mathcal{E}}_{\perp,ij} P_i \left( E_y L_x - E_x L_y \right) P_j \\
  & + \mathcal{E}_{\parallel,ij} E_z P_i \left( \proj{0} + \sum_{\sigma = \pm} \op{\sigma_i}{\sigma_j} \right) P_j
    \end{aligned}
\end{equation}
between the orbital spaces or inside the orbital singlet (i.e., for $i,j=1,2,3$ with $i \neq j$  or $i = j = 3$).
Here \(\Theta(x) \) is the Heaviside function which is $1$ for positive $x$ and else $0$.
Due to the different properties under time inversion, the matrix elements of $L_x$ and $L_y$ within an orbital doublet vanish, while $x$ and $y$ can be non-zero, such that the coupling to electric fields inside the orbital doublets ($i=1,2$) takes the form
\begin{align}
    \label{eq:VelDoublet}
    P_i V_{\mathrm{el}} P_i = \ & \mathcal{E}_{\parallel,ii} E_z \mathbb{I} + \mathcal{E}_{\perp,ii} \left( E_x \sigma_x - E_y \sigma_y \right),
\end{align}
where $\mathbb{I}$ is the identity matrix and $\sigma_x$ and $\sigma_y$ are the $x$ and $y$ Pauli matrices acting between $\ket{\pm_i}$.
In analogy to the Zeeman and spin-orbit terms, we combine $\mathcal{E}_{\perp,ij} = \tilde{\mathcal{E}}_{\perp,ij} l_{\perp,ij}$ and the factors $\mathcal{E}_{k,ij} = \mathcal{E}_{k,ji}$ are symmetric.
The expressions for the energies and Rabi frequencies in the following will only depend on the parameters $\epsilon_i, \lambda_{k,ij}, r_{k,ij}$ and $\mathcal{E}_{k,ij}$.
Using $C_{3v} \subset T_d$ and the angle $\phi$ it is possible to restrict \emph{parts} of the Hamiltonian to intact $T_d$ symmetry, see Appendix~\ref{sec:LTd} for more details.

\section{Results\label{sec:results}}
In the following we will study how the spin-orbit coupling influences the defect spin system and as such the interaction with external fields.
Comparing Eqs.~\eqref{eq:Vel} and \eqref{eq:Hz} we note that the orbital part of the Zeeman Hamiltonian  $H_{z}$ has a form which is similar to the coupling to external electric fields $V_{\mathrm{el}}$ between different orbital spaces.
As the Zeeman term additionally includes pure spin transitions we will concentrate on this term in the following, calculations for $V_{\mathrm{el}}$ are analogous and corresponding results will be summarized further below.

The Hamiltonian $H_{o} + H_{\mathrm{so}}$ can be block diagonalized into
two Hermitian $2\times2$ blocks and two real and symmetric $3\times3$ blocks.
The blocks of the same size are related to each other by time reversal symmetry.
This implies that there are at most 5 doubly degenerate eigenvalues, corresponding to the KDs.
The eigenstates of the $2\times2$ ($3\times3$) blocks transform according to $\Gamma_{5/6}$ ($\Gamma_4$).
The resulting eigenvalues are listed in Appendix~\ref{sec:AnSO}.

\subsection{Perturbation Theory\label{sec:SW}}
\begin{figure*}
\includegraphics[width=\linewidth]{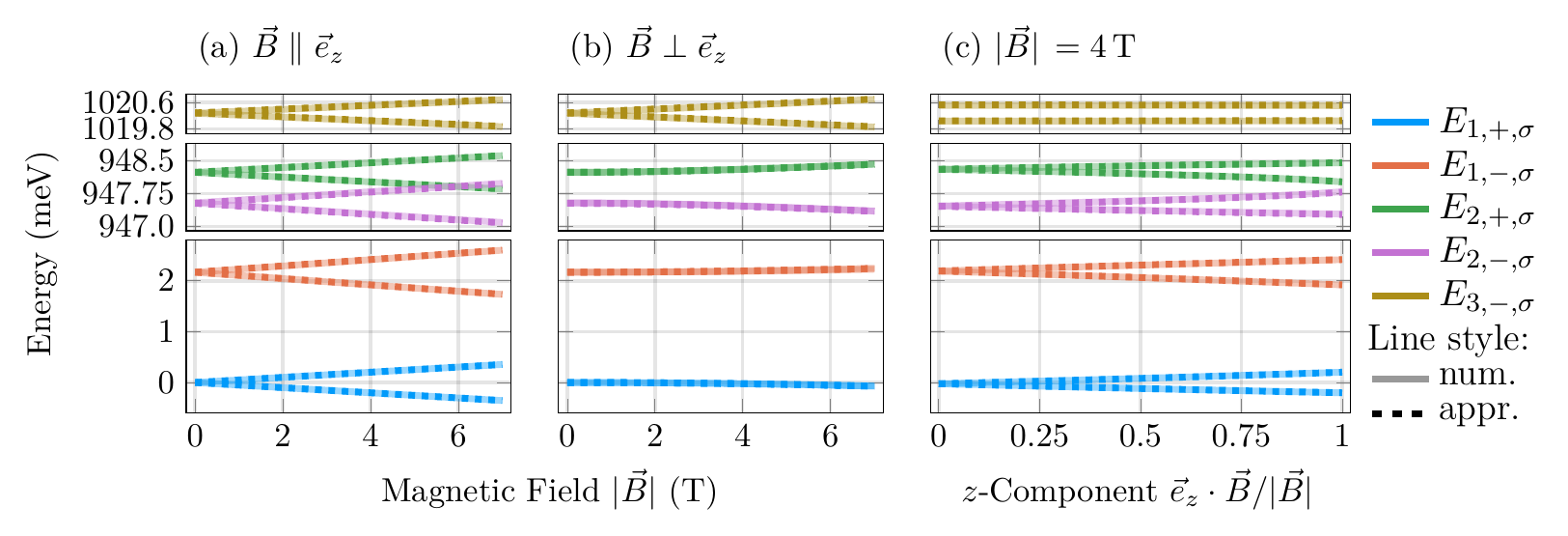}
\caption{\label{fig:Elvls}Comparison of numerically calculated level structure (faded solid) to the approximate levels calculated perturbatively in Eq.~\eqref{eq:Effx} (dashed) for  V in the $\alpha$-configuration of $6H$-SiC according to parameters used in~\cite{kaufmann97} (see Appendix \ref{sec:paramVSih6H}).
  In (a) [(b)] the dependence for a magnetic field purely in $z$[$x$]-direction is shown.
  The dependence on the direction for a fixed magnetic field strength is depicted in (c).
  We choose the energy scale such that the ground state spin-orbit energy $E_{1,+,\downarrow} = 0$ in the absence of a magnetic field ($\vec{B} = 0$).
  We observe a predominantly linear dependence of the energies on the $z$-component of the field as well as the higher order correction due to the $x$-component.}
\end{figure*}
Starting from the atomic energy levels, the largest contribution to the energy splitting is due to the crystal potential and the spin-orbit coupling is small compared to the resulting level spacings, $\lambda_{k,ij} \ll \epsilon_i-\epsilon_j$ for $i \neq j$.
%
Experiments and ab initio calculations~\cite{kaufmann97,csore19} show that the reduction factor of the spin-orbit coupling $\lambda_{k,ij}/\lambda_0$ and the orbital reduction of the Zeeman term $r_{k',ij}$ share the same order of magnitude and
hence, $\abs{\vec{B}} \ll B_{\mathrm{so}}$ implies $\mu_B r_{k',ij} \abs{\vec{B}} \ll \lambda_{k,ij}$.
We use a Schrieffer-Wolff transformation~\cite{bravyi11} treating the \emph{off-diagonal} elements of $H_{\mathrm{so}}$ as the perturbation
\begin{align}
  \label{eq:SWT}
      H_0 = & H_{o} + \sum_i P_i H_{\mathrm{so}} P_i,  & V = \sum_{i \ne j} P_i H_{\mathrm{so}} P_j
\end{align}
such that we find the effective Hamiltonian
\begin{equation}
  \label{eq:Heff}
  \begin{aligned}
    H_{\mathrm{eff}} = & \sum_i P_i e^S H e^{-S} P_i \\
    \approx & \sum_i P_i (H_0 + \frac{1}{2} \comm{S}{V} + H_{z} + \comm{S}{H_{z}}) P_i,
  \end{aligned}
\end{equation}
 for $\comm{H_0}{S} = V + \mathcal{O}[\lambda_{k,ij} \lambda_{k',ij'} / (\epsilon_i - \epsilon_j)]$ and
we enforce the additional constraint for  $P_i\comm{S}{H_{z}}P_i$ to be diagonal for $\vec{B} \parallel \vec{e}_z$.

These conditions ensure that $H_{\mathrm{eff}}$ lies inside the projected spaces up to second order in the spin-orbit coupling, i.e., $\lambda_{k,ij}^2/(\epsilon_i - \epsilon_j)$,
and is diagonal for $\vec{B} \parallel \vec{e}_z$.
Therefore, the diagonal elements correspond to the second order eigenvalues for a magnetic field parallel to the crystal axis, given by
\begin{widetext}
\begin{align}
  \label{eq:EnergieO2}
  \left. \begin{matrix} E_{i,\Gamma_{5/6}, \sigma}^{(2)} \\ E_{i,\Gamma_4, \sigma}^{(2)} \end{matrix} \right\}
  = \ &
  \epsilon_i
      \pm \frac{1}{2} \lambda_{\parallel,ii}
  + \frac{(-1)^i \lambda_{\parallel12}^{2}}{4 \left(\epsilon_2 - \epsilon_1\right)}
      + \left\{ \begin{matrix} - \frac{(-1)^i \lambda_{\perp12}^2}{2 \left(\epsilon_2 - \epsilon_1\right)}
        + \mu_B \frac{\sigma}{2} B_{z} g_{i,\Gamma_{5/6},\parallel} \\
        - \frac{1}{2} \frac{\lambda_{\perp,i3}^{2}}{\epsilon_3 - \epsilon_i}
        + \mu_B \frac{\sigma}{2} B_{z} g_{i,\Gamma_{4},\parallel}
      \end{matrix} \right. , \\
  E_{3, \Gamma_4, \sigma}^{(2)}
  = \ & \epsilon_3 + \frac{\lambda_{\perp,13}^{2}}{\epsilon_3 - \epsilon_1} + \frac{\lambda_{\perp,23}^{2}}{\epsilon_3 - \epsilon_2} + \sigma \frac{g_s}{2} \mu_{B} B_{z}, 
\end{align}
The corresponding first order eigenstates are given by the columns of $\I - S$,
\begin{align}
  \ket{i,\Gamma_{5/6},\sigma}^{(1)} = & \ket{\sigma_i} \ket{\sigma} + \sigma \frac{2 \lambda_{\perp,12} r_{\parallel,12}}{(\epsilon_2-\epsilon_1)(g_s + 2 r_{\parallel,ii})} \ket{-\sigma_i} \ket{-\sigma} - \frac{(-1)^i \lambda_{\parallel,12}}{2 \left( \epsilon_2-\epsilon_1 \right)} \ket{\sigma_{3-i}} \ket{\sigma} - \frac{\sigma \lambda_{\perp,12}}{\epsilon_2-\epsilon_1} \ket{-\sigma_{3-i}} \ket{-\sigma},  \label{eq:statesO2a}\\
  \ket{i,\Gamma_{4},\sigma}^{(1)} = & \ket{-\sigma_i} \ket{\sigma} - \frac{(-1)^i \lambda_{\parallel,12}}{2 \left( \epsilon_2-\epsilon_1 \right)} \ket{-\sigma_{3-i}} \ket{\sigma} - \frac{\sigma \lambda_{\perp,i3}}{\epsilon_3-\epsilon_i} \ket{0} \ket{-\sigma} \text{ for } i = 1,2, \text{ and } \label{eq:statesO2b} \\
  \ket{3,\Gamma_{4},\sigma}^{(1)} = & \ket{0} \ket{\sigma}+ \frac{\sigma \lambda_{\perp,13}}{\epsilon_3-\epsilon_1} \ket{\sigma_1} \ket{-\sigma} + \frac{\sigma \lambda_{\perp,23}}{\epsilon_3-\epsilon_2} \ket{\sigma_2} \ket{-\sigma},
  \label{eq:statesO2c}
\end{align}
\end{widetext}
with $i=1,2$, the parallel $g$-tensor components
\begin{align}
  \label{eq:gpar}
  \left. \begin{matrix} g_{i,\Gamma_{5/6},\parallel} \\ g_{i,\Gamma_{4},\parallel} \end{matrix} \right\}
  =\, & g_s \pm 2 r_{\parallel,ii} + (-1)^i 2 \frac{\lambda_{\parallel,12} r_{\parallel,12}}{\epsilon_2 - \epsilon_1},
\end{align}
and where we use $\sigma$ inside (outside) the spin ket to represent $\uparrow$ ($+$) and $\downarrow$ ($-$).
The first terms on the right hand side of the states Eqs.~(\ref{eq:statesO2a})-(\ref{eq:statesO2c}) are the reordered unperturbed states $\ket{i, \Gamma_{\gamma}, \sigma}^{(0)}$.
Without loss of generality, due to the symmetry of the Hamiltonian, we choose the coordinate system such that $\vec{B}$ lies in the $x,z$-plane.
Without an external magnetic field, these states approximate the zero-field analytic solutions and share their most important features.  At finite magnetic fields, Eq.~\eqref{eq:Heff} and Eqs.~(\ref{eq:statesO2a})-(\ref{eq:statesO2c}) provide a simpler description compared to an analytic solution.

For the approximate matrix elements of an operator $O$, we introduce the notation
\begin{align}
  \label{eq:secmel}
  \braket{i,\Gamma_{\gamma},\sigma|O|j,\Gamma_{\gamma},\sigma}_{\mathrm{eff}}
  \equiv&\, \braketo{i,\Gamma_{\gamma},\sigma|O_{\mathrm{eff}}|j,\Gamma_{\gamma},\sigma}{(0)},
\end{align}
where $O_{\mathrm{eff}} = O + \comm{O}{S}$.
Because the operators $O_{\mathrm{eff}}$ are of combined second order in the matrix elements of $O$ and $H_{\mathrm{so}}$ and in analogy to $H_{\mathrm{eff}}$ being second order in the spin-orbit coupling we will refer to them as the second order matrix elements.
Furthermore, we will refer to the elements of $O$ (between eigenstates of $H_0$) as first order matrix elements.

As a static magnetic field along the crystal axis only breaks time-reversal symmetry but not the $C_{3v}$ point symmetry, it can only mix states transforming according to the same irrep and lift the degeneracy of the KDs.
For $B_x \neq 0$ the effective Hamiltonians of the orbital doublets and the singlet are not diagonal.
The matrix elements inside the KDs are given by
$\braket{i,\Gamma_{k},\sigma | H | i, \Gamma_{k}, \sigma}_{\mathrm{eff}} = E_{i,\Gamma_{k},\sigma}^{(2)}$
and
\begin{align}
  \label{eq:Heffii}
  \braket{i,\Gamma_{4},\sigma | H | i, \Gamma_{4}, -\sigma}_{\mathrm{eff}} = & \mu_B g_{i,\perp} B_{x}/2,
\end{align}
where we define the perpendicular $g$-factors
\begin{align}
  \label{eq:Heff3}
  g_{i,\perp} = \begin{cases} g_{s} + \frac{4 \lambda_{\perp,13} r_{\perp,13}}{\epsilon_3 - \epsilon_i} + \frac{4 \lambda_{\perp,23} r_{\perp23}}{\epsilon_3 - \epsilon_2} \text{ for } i = 3 \\  - \frac{4 \lambda_{\perp,i3} r_{\perp,i3}}{\epsilon_3-\epsilon_i} \text{ for } i=1,2 \end{cases}.
\end{align}
We see that only the singlet KD has a non-suppressed effective perpendicular $g$-factor.
Agreeing with pure group theoretical considerations we find that the perpendicular $g$-factors for the $\Gamma_{5/6}$ KDs are zero.

The matrix elements mixing the KDs inside an orbital doublet are given by
\begin{align}
  \label{eq:Heff12}
  \braket{i,\Gamma_{5/6},\sigma | H | i, \Gamma_{4}, \sigma}_{\mathrm{eff}} = \, & \sigma \mu_B g_{i,c}  B_x/2 , \\
  \braket{i,\Gamma_{5/6},\sigma | H | i, \Gamma_{4}, -\sigma}_{\mathrm{eff}} = \, & \mu_B g_{i,f} B_x/2,
\end{align}
with the spin conserving ($c$) and flipping ($f$) mixing $g$-factors
\begin{align}
  \label{eq:gicf}
  g_{i, c} = & \frac{2 \lambda_{\parallel,12}r_{\perp,12}}{\epsilon_2 - \epsilon_1} +
  \frac{\lambda_{\perp,12} r_{\parallel,12}}{\epsilon_2 - \epsilon_1} \frac{2 g_s}{g_s + 2 r_{\parallel,ii}} , \\
  g_{i,f} = & g_s + (-1)^i 2 \left( \frac{\lambda_{\perp,12}r_{\perp,12}}{\epsilon_2 - \epsilon_1} + \frac{\lambda_{\perp,i3}r_{\perp,i3}}{\epsilon_3 - \epsilon_i} \right).
\end{align}
The only off-diagonal $g$-factors that do not vanish without spin-orbit coupling are $g_{i,f}$ with $i=1,2$ and $g_{4,\perp}$.
%
The $g$-factors as expressed above and evaluated for V in the $\alpha$-configuration of $6H$-SiC are within the margins of error for previously experimentally determined values~\cite{kaufmann97} when using the provided reduction factors according to parameters used in~\cite{kaufmann97} (see Appendix~\ref{sec:paramVSih6H}).
The maximal deviation for values that were assumed to be $0$ ($g_{i,\perp}$ and $g_{i,c}$ for $i=1,2$) is approximately $0.08$.
Previous derivations of the $g$-factors (see Appendix~A of Kaufmann \textit{et al.}~\cite{kaufmann97}) for a pair of $\Gamma_4$ and $\Gamma_{5/6}$ states  do not take the interaction with the remaining levels into account.
The effective Hamiltonian \eqref{eq:Heff} connects  the previous Hamiltonians to the Hamiltonian for the whole $d$ orbital.

For $T_d$ symmetry $\ket{i,\Gamma_{5/6},\sigma}$ can be paired with $\ket{i,\Gamma_4,-\sigma}$ as without an external field all four of these states are degenerate.
This shows that for pure $T_d$ symmetry, the effective spin levels have a non-suppressed perpendicular $g$-factor but are doubly degenerate (to first order).
This suggests that when the perpendicular $g$-factor vanishes for some states of such a system, the $C_{3v}$ breaking part of the Hamiltonian cannot be neglected.

The dominating effect of a static magnetic field in arbitrary direction can be taken into account
by diagonalizing the first order terms, i.e., the terms proportional to $g_s$.
This leads to the pseudo spin states (PSS),
\begin{align}
  \ket{i,+,\sigma} = 
  & \cos\theta_{i,\sigma} \ket{i,\Gamma_{5/6},\sigma} 
  + \sin\theta_{i,\sigma} \ket{i,\Gamma_{4},-\sigma}, \label{eq:statesO2pm} \\
  \ket{i,-,\sigma} =
  & \sin\theta_{i,-\sigma} \ket{i,\Gamma_{5/6},-\sigma} 
  - \cos\theta_{i,-\sigma} \ket{i,\Gamma_{4},\sigma},  \label{eq:statesO2pm2} \\
  \ket{3,\sigma} = & \cos\theta_{3} \ket{3,\Gamma_4,\sigma}_z + \sigma \sin\theta_3 \ket{2,\Gamma_{4},-\sigma},\label{eq:statesO2pm3}
\end{align}
with $i=1,2$, and mixing angles given by
\begin{align}
  \label{eq:anglepm}
 & \tan 2 \theta_{i,\sigma}  =  \frac{\mu_B g_s B_x}{\Delta_i+ \mu_B g_s \sigma B_z}, &
  \tan 2\theta_{3}  = \frac{B_x}{B_z} ,
\end{align}
in terms of the second order spin-orbit energy splittings
\begin{equation}
  \label{eq:Deltai}
  \begin{aligned}
  \Delta_i & = \lambda_{\parallel,ii} +
      \frac{\lambda_{\perp,i3}^{2}}{\epsilon_3 - \epsilon_i} +
      \frac{(-1)^i\lambda_{\perp12}^{2}}{\epsilon_2 - \epsilon_1},
  \end{aligned}
\end{equation}
which is given by the energy difference $E_{i,\Gamma_{5/6},\sigma} - E_{i,\Gamma_{4},\sigma}$ for $\vec{B}=0$ and is indicated in Fig.~\ref{fig:Ediag}.
The PSS are not related by time-reversal symmetry and can therefore be coupled by electric fields.

Neglecting $\comm{S}{H_{z}}$, i.e., terms proportional to $r_{k',ij'} \lambda_{k,ij}/(\epsilon_i - \epsilon_j)$, the diagonal entries in the basis consisting of the PSS correspond to approximate eigenvalues
\begin{widetext}
\begin{align}
  \label{eq:Effx}
  E_{i,\pm,\sigma}^{(2)} & 
   = \braketo{i,\pm,\sigma| H_{\mathrm{eff}} | i,\pm,\sigma}{(0)} \notag \\
   = & \, 
   \epsilon_i
      - \frac{\lambda_{\parallel12}^{2} + 2 \lambda_{\perp12}^2}{4 \left(\epsilon_2 - \epsilon_1\right)}
      - \frac{\lambda_{\perp,i3}^{2}}{2 \epsilon_3 - 2 \epsilon_1}
      \pm \sigma B_{z} \mu_B r_{\parallel,ii} \pm \frac{\operatorname{sign}(\Delta_i + \mu_B g_s \sigma B_z)}{2} \sqrt{\left[ \Delta_i + \mu_B g_s \sigma B_z \right]^2 + \left( \mu_B g_s B_x \right)^2 }, \\
  E_{3,\sigma}^{(2)} & = \braketo{3,\sigma| H_{\mathrm{eff}} | 3,\sigma}{(0)}
  = \epsilon_3 + \frac{\lambda_{\perp,13}^{2}}{\epsilon_3 - \epsilon_1} + \frac{\lambda_{\perp,23}^{2}}{\epsilon_3 - \epsilon_2} + \sigma \frac{g_s}{2} \mu_{B} \sqrt{B_{z}^2 + B_{x}^2} ,
\end{align}
\end{widetext}
as shown in Fig.~\ref{fig:Elvls}.
The states are labeled according to the crystal field eigenspace $i$, the spin-orbit splitting $\pm$ such that for $B_x=0$ ``$+$'' (``$-$'') coincides with the states transforming according to $\Gamma_{5/6}$ ($\Gamma_4$).
These labels are independent of the level ordering, reflecting that the level ordering depends on the precise configuration of the defect, i.e. the free and \emph{a priori} unknown parameters of our model.
The results in the following do not depend on the level order.
Lastly $\sigma$ labels the effective spin.

The approximate eigenvalues $E_{i,\pm,\sigma}^{(2)}$ already yield good results for the energies for V in the $\alpha$-configuration of $6H$-SiC compared to the numerical model by Kaufmann \textit{et al.}~\citep{kaufmann97}, see Fig.~\ref{fig:Elvls}.
For a magnetic field in $z$ direction all the states split linearly [Fig.~\ref{fig:Elvls}(a)].
For magnetic fields along the $x$-axis only the $i=3$ level splits linearly while the remaining PSS stay degenerate in the first order, the KDs of one orbital doublet are pushed further apart with increasing magnetic field in $x$-direction [Fig.~\ref{fig:Elvls}(b)].
For a small fixed magnetic field strength we see an approximate linear dependence of the splitting of the KD as a function of the $z$-projection of the magnetic field for the states of the orbital doublets and for the singlet PSS the dependence is approximately constant [Fig.~\ref{fig:Elvls}(c)].

If we consider $V_{\mathrm{el}}$ instead of $H_{z}$ the biggest difference is that we do not have the pure spin matrix elements ($\propto g_s$) but instead have an allowed spin-conserving coupling between the KDs in the same orbital doublet. 
Furthermore, the KDs stay degenerate because $V_{\mathrm{el}}$ does not break the time-reversal symmetry for static electric fields.
The $E_z$ fields action inside the projected spaces is given by
\begin{align}
  & \left. \begin{matrix} \braket{ i, \Gamma_{5/6}, \sigma | V_{\mathrm{el}} | i, \Gamma_{5/6}, \sigma } \\ \braket{ i, \Gamma_{4}, \sigma | V_{\mathrm{el}} | i, \Gamma_{4}, \sigma } \end{matrix} \right\} 
  = \Bigg[ \mathcal{E}_{\parallel, ii}
  \pm (-1)^i \frac{\mathcal{E}_{\parallel, 12} \lambda_{\parallel, 12}}{\epsilon_2 - \epsilon_1} \Bigg] E_{z},
  \label{eq:EzO2}
\end{align}
for $i=  1,2$  and $\braket{3, \Gamma_4, \sigma| V_{\mathrm{el}} | 3, \Gamma_4, \sigma } = \mathcal{E}_{\parallel, 33} E_z$.

\subsection{Magnetic (and Optical) Resonance Properties \label{sec:resonance}}
\begin{figure*}
\includegraphics[width=\linewidth]{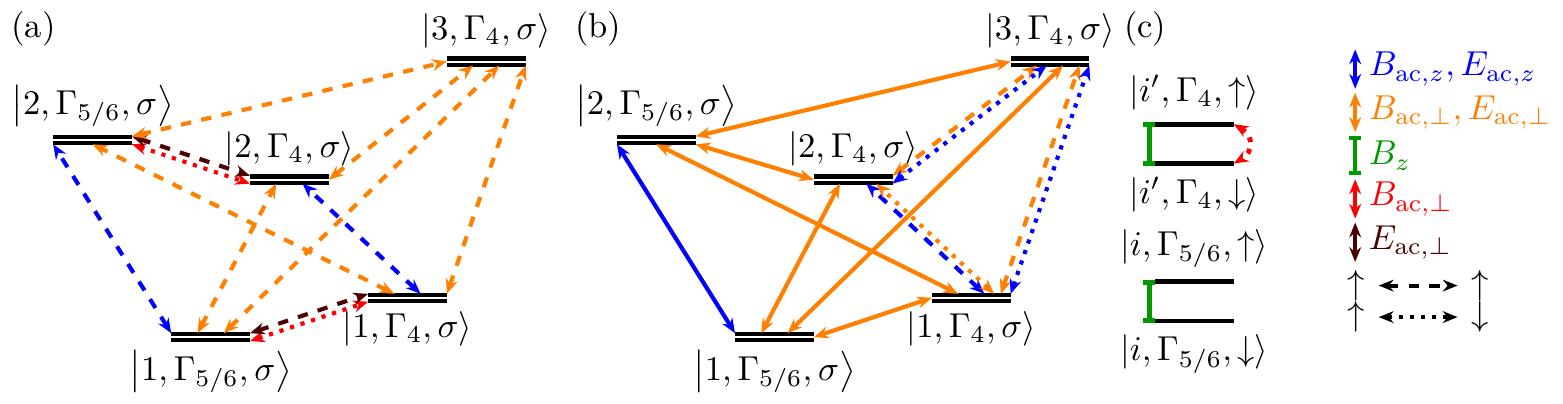}
\caption{\label{fig:Etrans}
   Non-zero matrix elements of the driving Hamiltonian $H_{d}$ between KD states for  $\vec{B}_0\parallel\vec{e}_z$ taking
  (a) only reordering of the states due to spin-orbit coupling
  $^{(0)}\hspace{-0.2em}\braket{i,\Gamma_{\gamma},\sigma | H_{d} | i,\Gamma_{\gamma'},\sigma'}^{(0)}$ (first order in $H_{d}$) and
  (b) the correction of the states due to the spin-orbit coupling $\braket{i,\Gamma_{\gamma},\sigma | H_{d} | i,\Gamma_{\gamma'},\sigma'}_{\mathrm{eff}}$ (mixed second order in $H_{d}$ and $H_{\mathrm{so}}$) into account.
  In (c) the mixed second order matrix elements inside the KDs are depicted with the splitting due to the static magnetic field along the crystal axis (green) that varies due to $B_{\mathrm{ac},z}$.
  Dashed lines connect states of the same spin $\ket{i,\Gamma_{\gamma},\sigma} \leftrightarrow \ket{j,\Gamma_{\gamma'},\sigma}$, dotted lines states of inverted spin $\ket{i,\Gamma_{\gamma},\sigma} \leftrightarrow \ket{j,\Gamma_{\gamma'},-\sigma}$ and solid lines show that both channels are possible.
  The color indicates which field component gives rise to the transition (see legend) where $\perp$ corresponds to a field in the $x,y$-plane.
  The red line connecting inverted spin states of the different irreps of the same orbital doublet corresponds to a pure spin transition, therefore in the first order it can only be driven by a magnetic field.
  In contrast, the corresponding spin conserving transition can be
  driven only by an electric field in the first order.
}
\end{figure*}
As an electrical field cannot lift the degeneracy of a KD we will consider the case where we apply a static magnetic field $\vec{B}_0$ that splits the KD and a periodical (electric or magnetic field) driving, i.e., of the form
$\vec{B}_{\mathrm{ac}} \sin(\omega t)$ or $\vec{E}_{\mathrm{ac}} \sin(\omega t)$.
The driving field gives rise to an additional part of the Hamiltonian
$H_{d} = O \sin(\omega t)$, 
where $O$ denotes either $H_{z}$ with the driving magnetic field amplitude $\vec{B} = \vec{B}_{\mathrm{ac}}$
or the electric driving term $V_{\mathrm{el}}$ with $\vec{E} = \vec{E}_{\mathrm{ac}}$.
%
As before we will discuss the magnetic field case in detail and summarize the main differences for the electric field.

For simplicity, we define
\begin{equation}
  \label{eq:RRFreq}
  \begin{aligned}
  \Omega_{\Psi \Psi'} & = \frac{1}{2 \hbar} \abs{\braket{\Psi|H_{d}|\Psi'}}.
  \end{aligned}
\end{equation}
For $\Psi, \Psi'$ eigenstates of $H_{\mathrm{cr}} + H_{\mathrm{so}} + H_{z}\Big|_{\vec{B} = \vec{B}_0}$ and $\Psi \neq \Psi'$ $\Omega_{\Psi \Psi'}$ is the resonant Rabi frequency.
We approximate $\Omega_{\Psi \Psi'} \approx \Omega_{\Psi \Psi'}^{\mathrm{eff}} =  \abs{\braket{\Psi|H_{d}|\Psi'}_{\mathrm{eff}}}/ {2 \hbar}$, using the notation of Eq.~\eqref{eq:secmel} but where $\Psi$ and $\Psi'$ are eigenstates of $H_{\mathrm{eff}}$,
thereby taking state mixing due to field components perpendicular to the crystal axis into account.

\subsubsection{Static Magnetic Field along Crystal Axis}
\begin{figure}
\includegraphics[width=\linewidth]{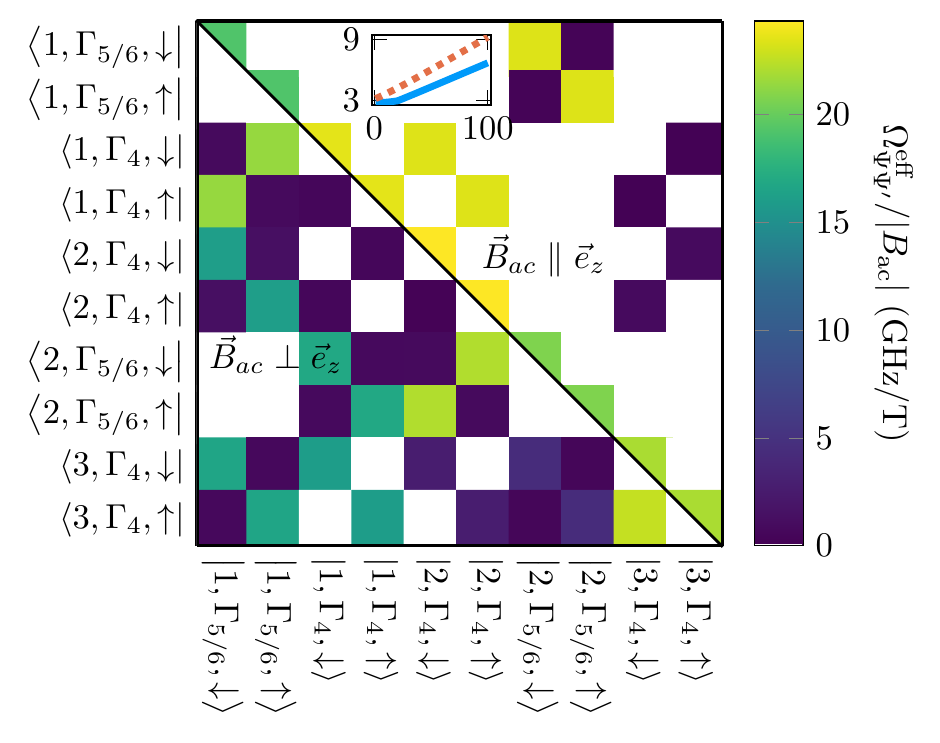}
\caption{\label{fig:EtransQuant}
Rabi frequencies on resonance in second order (off-diagonal matrix elements) and energy level oscillation amplitudes (diagonal)
$\Omega^{\mathrm{eff}}_{\Psi\Psi'}/\abs{B_{\mathrm{ac}}}$ according to Eq.~\eqref{eq:RRFreq}.  Parameters are chosen for V in the $\alpha$-configuration of $6H$-SiC according to parameters used in~\cite{kaufmann97} (see Appendix~\ref{sec:paramVSih6H}) and a magnetic driving field.
  The static magnetic field $B_0 \vec{e}_z$ is parallel to the crystal axis and the oscillatory magnetic field perpendicular (bottom left half) or parallel (top right half).
  The $x$- and $y$-axes respectively label the states $\Psi$ and $\Psi'$.
  The inset depicts the relative error in units of $10^{-3}$ according to Eq.~\eqref{eq:errRFO2} as a function of $B_0$ ($x$-axis) for $\vec{B}_{\mathrm{ac}} \parallel \vec{e}_x$ (solid blue line) and  $\vec{B}_{\mathrm{ac}} \parallel \vec{e}_z$ (orange dotted line).
  We plot matrix elements that identically vanish in white.
}
\end{figure}

We  consider a static field aligned with the crystal axis $\vec{B}_0 \parallel \vec{e}_z$; in this case we can without loss of generality consider the oscillatory field to lie in the $x,z$-plane.
Also, time-reversal symmetry is broken by $\vec{B}_0$ while the point symmetry of the defect stays intact, implying that the PSSs coincide with the KDs and the second order $\Omega^{\mathrm{eff}}$ is independent of $\abs{\vec{B}_0}$.

The matrix elements of $H_{d}$ for a static magnetic field along the $z$-axis depend only on the field component parallel \emph{or} perpendicular to the crystal axis (but not both).
The resulting structure within first order is sketched in Fig.~\ref{fig:Etrans}(a).
We start by only taking the first order effect of the spin-orbit coupling into account, i.e., the reordering of the unperturbed states.
This corresponds to evaluating the matrix elements
$\braketo{i,\Gamma_{\gamma},\sigma|H_{d}|j,\Gamma_{\gamma'},\sigma'}{(0)}$.
The only non-zero matrix elements proportional to the (electric or magnetic) field parallel to the crystal axis are between states of the same irrep with the same spin ($\gamma=\gamma'$ and $\sigma=\sigma'$).
All other non-zero matrix elements are proportional to perpendicular field components.
Inside the orbital doublets ($i = j$) the elements connecting inverted spins ($\sigma = -\sigma'$) of different irreps ($\gamma \neq \gamma'$) are linked by perpendicular magnetic fields, while electric fields can drive the spin conserving ($\sigma = \sigma'$) transition.
Driving with a transverse field couples states with the same spin ($\sigma = \sigma'$) of different irreps ($\gamma \neq \gamma'$) from different orbital doublets ($i \neq j$) as well as states with the same spin between the singlet $\Gamma_4$ and all other irreps.
Inside KDs the only first-order non-diagonal matrix element is between the $\ket{3,\Gamma_4,\sigma}$ states and proportional to $B_{\mathrm{ac},x}$; electric fields cannot drive transitions inside KDs.

The mixing of the states due to the spin-orbit coupling, see Eqs.~(\ref{eq:statesO2a}) and (\ref{eq:statesO2b}) for the perturbative eigenstates, allows additional transitions.
The allowed transitions within second order perturbation theory are shown in Fig.~\ref{fig:Etrans}(b).
Summarizing the spin-orbit coupling mixes all the $\Gamma_{5/6}$ (spin) states with each other while
$\Gamma_4$ states of the orbital doublets with the same spin are mixed with each other and with states of the singlet of inverted spin.
This leads to all first order allowed transitions between a basis state of the $\Gamma_{5/6}$ irrep being allowed conserving \emph{and} flipping the spin in the second order.
Additionally, to the first order transitions, transitions between states that both transform like basis states of $\Gamma_4$ spin flipping transitions are allowed inside the KDs for $B_{\mathrm{ac,x}}$ and between the states of the orbital doublets for $B_{\mathrm{ac},x}$ and $E_{\mathrm{ac},x}$ in the second order.
Between the singlet and the orbital doublet states transforming according to $\Gamma_4$ inverted spin states couple proportional to $B_{\mathrm{ac},z}$ and $E_{\mathrm{ac},z}$ in the second order.
The structure given by the second order coincides with the structure for the analytic spin-orbit eigenstates.
The matrix elements $\Omega_{\Psi\Psi'}^{\mathrm{eff}}/\abs{B_{\mathrm{ac}}}$ for V in the $\alpha$-configuration of $6H$-SiC are depicted in Fig.~\ref{fig:EtransQuant}.
The maximum relative error of the $\Omega_{\Psi \Psi'}$ as a function of $B_0 \vec{e}_z$ and $\vec{B_{\mathrm{ac}}}$ is given by
\begin{align}
  \label{eq:errRFO2}
  \max_{\Phi}(\abs{\Omega^{\mathrm{eff}}_{\Phi \Phi} - \Omega_{\Phi \Phi}})/\max_{\Phi}(\Omega_{\Phi \Phi}),
\end{align}
where $\Phi$ is an arbitrary wavefunction.
The error is smaller than $1\%$ for arbitrary driving magnetic field strength for V in the $\alpha$-configuration of $6H$-SiC according to parameters used in~\cite{kaufmann97} (see Appendix~\ref{sec:paramVSih6H}) for static magnetic field strength $B_0 < 100\,$T, as can be seen in the inset of Fig.~\ref{fig:EtransQuant}.

Considering that \emph{at least for} $V$ in the $\alpha$-configuration of $6H$-SiC the leading order is much larger than the following orders, it is hard to drive a transition that is suppressed in the leading order.
For example if the Zeeman splitting is much smaller than the crystal splitting it is difficult to drive the transition $\ket{1,\Gamma_{5/6},\downarrow} \leftrightarrow \ket{2,\Gamma_4,\uparrow}$ because instead one would drive the transition $\leftrightarrow \ket{2,\Gamma_4,\downarrow}$ off-resonantly.
This explains the missing measurement points in Figs.~6~and~7 in the paper by Kaufman \textit{et al.}~\cite{kaufmann97} along the $\alpha$-lines.

\subsubsection{Static Magnetic Field in Arbitrary Direction}
\begin{figure*}
\includegraphics{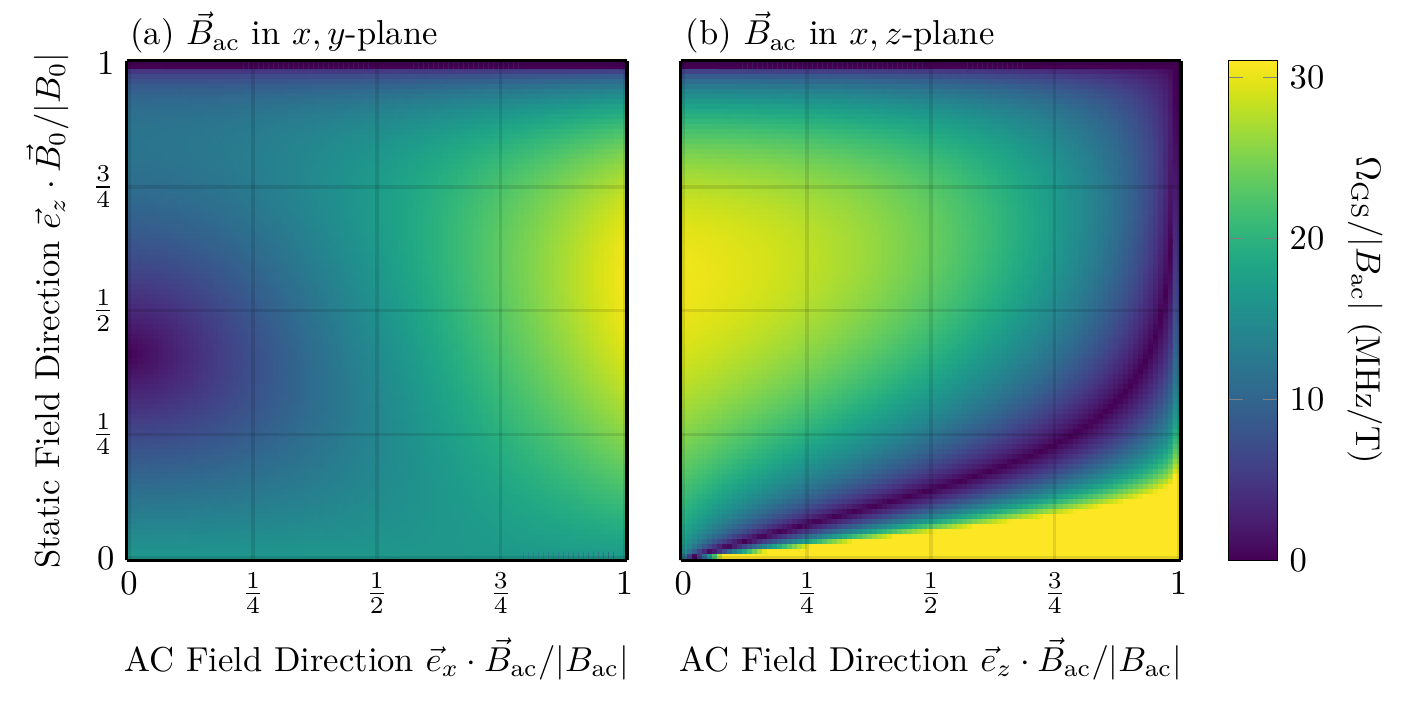}
\caption{\label{fig:GSRabiDir}Resonant Rabi frequency $\Omega_{\mathrm{GS}}/\abs{B_{\mathrm{ac}}}$ (MHz/T) between the ground state pseudo spin states as a function of the static and ac magnetic field direction for parameters according V in the $\alpha$-configuration of $6H$-SiC according to parameters used in~\cite{kaufmann97} (see Appendix~\ref{sec:paramVSih6H}) and a static magnetic field strength $B_0 = 3\,$T.
  In (a) the ac field lies in the $x,y$-plane and in (b) in the $x,z$-plane.
  The yellow area in the bottom right of (b) corresponds to values $\geq 31\,$MHz$/$T.
The approximate Rabi frequency $\Omega_{\mathrm{GS}}^{\mathrm{eff}}$ agrees quantitatively up to a systematic deviation of approximately $+15\%$ apart from the region where $\Omega_{\mathrm{GS}}/\abs{B_{\mathrm{ac}}}<1\,$MHz$/$T there the absolute error is smaller than $0.2\,$MHz$/$T.}
\end{figure*}
For $B_{0,x} \neq 0$ the point symmetry and time-reversal symmetry are broken by the static magnetic field.
This is already manifested in the first order effect of a static magnetic field in $x$-direction,
the mixing of states of inverted spin and different irreps inside the orbital doublets, see Eqs.~(\ref{eq:statesO2pm})-(\ref{eq:statesO2pm3}).

We first discuss the special case $B_{0,z} = 0$ where the PSSs fulfill $E_{i,\pm,\sigma}^{(2)} = E_{i,\pm,-\sigma}^{(2)}$ ($i=1,2$), see Eq.~\eqref{eq:Effx}.
When considering the states (not the energies) this makes the small second order contribution between the approximatly degenerate PSSs very relevant, leading to states that diagonalize the coupling of the driving magnetic field in $x$ direction inside the PSS doublet.
Therefore, driving with an oscillatory magnetic field in $x$-direction will be suppressed while driving using a magnetic field in the $y$- or $z$-direction is possible.
This underlines that in case of a static field not aligned with the crystal axis the effective driving Hamiltonian can be anisotropic in all oscillatory field components.
Furthermore, the mixing of states makes several transitions possible, e.g., the $\ket{1,+,\downarrow} \leftrightarrow \ket{1,+,\uparrow}$ transition for $B_{\mathrm{ac}, y}$, $E_{\mathrm{ac}, x}$ and $E_{\mathrm{ac}, y}$. 

Now we consider a magnetic field along an arbitrary direction on the $x,z$-plane.
As an example, we discuss the transition within the PSS doublet $\ket{1,+,\sigma}$,
which constitutes the GS for some Mo and V defect configurations~\cite{kunzer93,kaufmann97,bosma18,spindlberger19,wolfowicz20,gilardoni20,csore19}.
Group theoretical selection rules imply that the transition inside the KD cannot be driven for $\vec{B}_0 \parallel \vec{e}_z$~\cite{gilardoni20}.
However, due to the mixing of the KDs for $\vec{B}_0 \nparallel \vec{e}_z$ the transition becomes possible.

In Fig.~\ref{fig:GSRabiDir} we show the Rabi frequency
$\Omega_{\mathrm{GS}} = \Omega_{\ket{1,+,\downarrow}, \ket{1,+,\uparrow}}$
as a function of the directions of the static and alternating magnetic fields.
There is a maximum value for the transition for $B_{0,z}\neq 0$ along the line where $\vec{B}_{\mathrm{ac}} \parallel \vec{e}_x$.
Using a static magnetic field corresponding to this maximum ensures a splitting of the energies of the PSSs while at the same time making it possible to drive the transition with a magnetic field along the $x$-axis.

Analogously the mixing of the states due to a static magnetic field perpendicular to the crystal axis can make other transitions (including between different PSSs of different crystal eigenspaces) possible for magnetic as well as electric driving fields.
To understand the mixing quantitatively and to maximize the resonant Rabi frequency one can use the effective Hamiltonian \eqref{eq:Heff}.

\section{Conclusions\label{sec:conclusion}}
We have introduced a framework that describes how the interplay of the reduced symmetry of a defect implanted in a solid and spin-orbit coupling give rise to a non-trivial spin structure with the direct application to a spin-1/2 defect in SiC.
We derived analytic energy levels in absence of external magnetic fields
and used perturbation theory to obtain effective Hamiltonians inside the crystal eigenspaces that are directly related to the Hamiltonian of the full orbital subspace.

This effective Hamiltonian directly links the anisotropy of the $g$-tensor of the KDs to the interplay of $C_{3v}$ symmetry and the spin-orbit coupling and can be used to explain the expected magnetic and optical resonance properties of the system for static magnetic fields in arbitrary direction.
We were able to show how transitions that are forbidden for intact $C_{3v}$ symmetry can be accessed by applying a static field perpendicular to the crystal axis.
This does not conflict with previous selection rules as external fields not aligned with the crystal axis break the $C_{3v}$ point symmetry of the defect.
We expect the employed theory to be useful to study allowed transitions for optical and microwave control, as well as relaxation and coherence times, and to optimize static fields to achieve desired forms of the driving Hamiltonian.
Looking forward, the derived Schrieffer-Wolff transformation can be used to construct effective hyperfine Hamiltonians for the defect states originating from the atomic $d$ orbital.

To obtain more quantitative information on the properties of the defects at various crystal sites it would be of great interest to compute  complete sets of reduction factors for different defect configurations.
Furthermore, reduction factors for $V_{\mathrm{el}}$ and $H_{\mathrm{hf}}$ would make it possible to study the effect of these using the employed framework in more detail and lead to more precise predictions.

\begin{acknowledgments}
  We thank C. Gilardoni, C. H. van der Wal and M. Trupke for insightful discussions.
  We acknowledge funding from the EU H2020 FET project QuanTELCO (862721).
\end{acknowledgments}

\appendix

\section{Symmetry Groups of the Defect\label{sec:symgroups}}
Choosing the $z$-direction parallel to the crystal axis and using the vectors $\vec{a}_i$ to the nearest neighboring atoms of the defect atom, see Fig.~\ref{fig:sym}, the symmetry operations that leave the defect site invariant are the identity $E$, the rotations $R_{\pm}$ around $\vec{e}_z$ by $\pm 2 \pi/3$, and reflections $\sigma_i$ on planes spanned by $\vec{a}_4$ and $\vec{a}_i$, $i = 1,2,3 $.
These symmetry operations are the elements of the $C_{3v}$ point group~\cite{tinkham03}.
This leads to the irreps of $C_{3v}$ given by two one-dimensional irreps $A_1$, $A_2$ and one two-dimensional irrep $E$~\cite{koster63}.
The double group $\overline{C_{3v}}$ additionally has the two-dimensional spinor representation $\Gamma_4$ and the two one-dimensional irreducible representations $\Gamma_5$ and $\Gamma_6$.

If we only consider the nearest neighboring atoms we have the symmetries given above as well as reflections on the planes spanned by $\vec{a}_i$ and $\vec{a}_j$, rotations by $\pm 2 \pi/3$ around $\vec{a}_i$, rotations by $2 \pi/2$ around $\vec{a}_i + \vec{a_4}$, and improper rotations by $2 \pi/4$ around $\vec{a}_i + \vec{a_4}$ where $i \neq j = 1,2,3 $.
These operations are the elements of the $T_d$ point group~\cite{tinkham03}.
$T_d$ has two one-dimensional, one two-dimensional and two three-dimensional irreps~\cite{koster63}.
The double group $\overline{T_d}$ additionally has two two-dimensional and one four-dimensional irrep.

\section{Transformations of States and Operators\label{sec:trafo}}

The representation of the rotation about an axis $\vec{n}$ with rotation angle $\alpha$
on angular momentum eigenstates is given by
\begin{align}
  \label{eq:rotationLM}
  {R}(\alpha, \vec{n}) \ket{l,m} = e^{-i \alpha \vec{n}\cdot {\vec{L}}} \ket{l,m}
\end{align}
and inversion is given by
\begin{align}
  \label{eq:inversionLM}
  {P} \ket{l,m} = (-1)^l \ket{l,m} .
\end{align}
With these considerations we can calculate the representation matrices of the $C_{3v}$ symmetry operators for the $\ket{m} = \ket{l=2, m}$ [see Eq.~\eqref{eq:base}] angular momentum states (a basis for the reducible representation $\Gamma_{l=2}$ of $C_{3v}$).
The resulting representations for the irreps are shown in Table~\ref{tab:symops}.
\begin{table}
  \caption{\label{tab:symops} Representation (matrices) of the symmetry operations of the $C_{3v}$ group for its irreps. The representation for $E$ is given for the basis states $\ket{\pm 1}$ and we use $\epsilon = e^{- \mathrm{i} 2 \pi / 3} $.}
  \begin{ruledtabular}
    \begin{tabular}{c|rrrrrr}
      irrep & $E$ & $R_+$ & $R_-$ & $\sigma_1$ & $\sigma_2$ & $\sigma_3$ \\
      $A_1$ & $1$ & $1$ & $1$  & $1$ & $1$ & $1$ \\
      $A_2$ & $1$ & $1$ & $1$  & $-1$ & $-1$ & $-1$ \\
      $E$ & $\I$ & $\begin{pmatrix} \epsilon & 0 \\ 0 & \epsilon^{*} \end{pmatrix}$ & $\begin{pmatrix}^{*} \epsilon & 0 \\ 0 & \epsilon \end{pmatrix}$ & $-\begin{pmatrix} 0 & 1 \\ 1 & 0 \end{pmatrix}$ & $-\begin{pmatrix} 0 & \epsilon^{*} \\ \epsilon & 0 \end{pmatrix}$ & $-\begin{pmatrix} 0 & \epsilon \\ \epsilon^{*} & 0 \end{pmatrix}$
    \end{tabular}
  \end{ruledtabular}
\end{table}

We find \( \ket{0} \) is a basis for irrep $A_1$.
The operators $H_{o}$, $z$, and $p_z$ all transform according to the irrep $A_1$.
The pairs of states \( (\ket{1}, \ket{-1}) \) and \( (-\ket{-2}, \ket{2}) \) are bases for the irrep \( E \).
The spherical components of the angular momentum operator $(L_{+1}, L_{-1})$ transform according to the irrep $E$ in the same way as the basis $(\ket{1}, \ket{-1})$, and are given by \( {L}_{\pm 1} = - i {({L}_{x} \pm i {L}_{y})}/{\sqrt{2}} = - i {{L}_{\pm}}/{\sqrt{2}} \).
The coordinate components $x,y$ also transform like $-L_y, L_x$ according to the irrep $E$.
The $z$-component of the angular momentum operator $L_z$ transforms according to the irrep $A_2$.
Representation matrices for $T_d$ symmetry operations can be calculated analogously but in this case the spherical harmonics cannot be assigned directly to irreps.

\section{Clebsch-Gordan Coefficients and the Wiegner-Eckart Theorem}
The Clebsch-Gordan coefficients link tensor-product states to basis states of an irrep~\cite{cornwell97}
\begin{align}
  \label{eq:wieckthm}
\ket{\xi^{r, \alpha}_l} = \sum_{j = 1}^{d_p} \sum_{k=1}^{d_q} \Big( \begin{matrix} p & q \\ j & k \end{matrix} \Big\vert \begin{matrix} r & \alpha \\ l & \end{matrix} \Big)
\ket{\phi^p_j} \otimes \ket{\psi^q_k},
\end{align}
where the state \( \ket{\psi^q_k}\ (\ket{\phi^p_j}, \ket{\xi_l^{r, \alpha}}) \) transforms according to row \( k\  (j, l) \) of irrep \( \Gamma^q\ (\Gamma^p, \Gamma^r) \).
The irrep \( \Gamma^r \) is contained $n_{pq}^r$ times in the product representation \( \Gamma^p \otimes \Gamma^{q} \) and $\alpha$ runs from $1$ to $n_{pq}^r$.
\begin{table}
  \caption{\label{tab:cgc} Clebsch-Gordan coefficients for $C_{3v}$ symmetry. Here $\pm$ correspond to basis states of $E$ transforming like $\ket{\pm 1}$.}
  \begin{ruledtabular}
    \begin{tabular}{c|c}
$\begin{aligned}
\Big( \begin{matrix} \Gamma_i & A_1 \\ k & 1 \end{matrix} \Big\vert \begin{matrix} \Gamma_j \\ m \end{matrix} \Big)
  = & \Big( \begin{matrix} A_1 & \Gamma_i \\ 1 & k \end{matrix} \Big\vert \begin{matrix} \Gamma_j \\ m & \end{matrix} \Big) = \delta_{ij} \delta_{km} \\
\Big( \begin{matrix} A_2 & A_2 \\ 1 & 1 \end{matrix} \Big\vert \begin{matrix} A_1 \\ 1 \end{matrix} \Big) = & 1 \\
\Big( \begin{matrix} E & A_2 \\ \pm & 1 \end{matrix} \Big\vert \begin{matrix} E \\ \pm \end{matrix} \Big)
  = & \Big( \begin{matrix} A_2 & E \\ 1  & \pm \end{matrix} \Big\vert \begin{matrix} E \\ \pm \end{matrix} \Big) = \pm 1 \\
\end{aligned}$ &
$\begin{aligned}
\Big( \begin{matrix} E & E \\ \pm & \mp \end{matrix} \Big\vert \begin{matrix} A_1 \\ 1 \end{matrix} \Big) = & \frac{1}{\sqrt{2}} \\
\Big( \begin{matrix} E & E \\ \pm & \mp \end{matrix} \Big\vert \begin{matrix} A_2 \\ 1 \end{matrix} \Big) = & \pm \frac{1}{\sqrt{2}} \\
\Big( \begin{matrix} E & E \\ \pm & \pm \end{matrix} \Big\vert \begin{matrix} E \\ \mp \end{matrix} \Big) = & \mp 1
\end{aligned}$
    \end{tabular}
  \end{ruledtabular}
\end{table}
Using the transformation matrices we can also calculate the Clebsch-Gordan coefficients for $C_{3v}$, see Table~\ref{tab:cgc}.
Equivalent coefficients for other bases of $E$ are found in~\cite{doherty13}.

To use the symmetry properties to derive the general form of the Hamiltonian we use the \emph{Wigner-Eckart theorem}~\cite{cornwell97}
\begin{align}
  \label{eq:WET}
\bra{\psi^r_l}{Q^q_k}\ket{\phi^p_j} = \sum_{\alpha = 1}^{n^r_{pq}} \Big( \begin{matrix} p & q \\ j & k \end{matrix} \Big\vert \begin{matrix} r & \alpha \\ l & \end{matrix} \Big)^* \braket{r || Q^q || p}_{\alpha},
\end{align}
where $Q_k^q$ is an operator transforming like a $k$ basis vector of irrep $\Gamma^q$,
$\braket{r || Q^q || p}_{\alpha}$ are reduced matrix elements and the Clebsch-Gordan coefficients are complex conjugated.
In this article we treat the reduced matrix elements as independent parameters that are given by experiments or ab initio calculations.

\section{Defect Hamiltonian and Crystal Potential\label{sec:HoSphHarm}}

The TM defect and crystal potentials only act on the orbital part of the wavefunction and
transform according to the irreducible representation $A_1$ of $C_{3v}$.
Furthermore, using time-reversal symmetry as well as Hermiticity of the Hamiltonian combining the defect atomic Hamiltonian with the crystal potential, we find
\begin{equation}
  \label{eq:HcrND}
  \begin{aligned}
  H_{o}
  = \ & H_{\mathrm{TM}} + H_{\mathrm{cr}}
  = \braket{A_1 \Vert H_{o} \Vert A_1} \proj{0} \\
  & + \braket{ E \Vert H_{o} \Vert E }_{12} \left( \op{-2}{+1} - \op{2}{-1} \right) \\
  & +  \sum_{i=1,2} \braket{ E \Vert H_{o} \Vert E }_{i} \left( \proj{+i} + \proj{-i} \right),
  \end{aligned}
\end{equation}
where all reduced matrix elements correspond to real parameters.
We introduce the definitions
\begin{align}
  \label{eq:HcrRMELs}
  \braket{A_1||H_{o}||A_1} & = \epsilon_3, \\
  \braket{E||H_{o}||E}_1 & = \epsilon_3 - {2 \Delta_c}/{3} + K'/2, \\
  \braket{E||H_{o}||E}_2 & = \epsilon_3 - {\Delta_c}/{3} + K'/2 - K, \text{ and } \\
  \braket{E||H_{o}||E}_{12} & = {\sqrt{2}\Delta_c}/{3},
\end{align}
where the reduced matrix elements are introduced in Eq.~\eqref{eq:WET}.
Here $\epsilon_3$ and $\Delta_c$ parameterize the part of the crystal Hamiltonian that fulfills the symmetry operations of $T_d$ and includes the off-diagonal elements in this basis.
Non-zero $K$ and $K'$ reduce the $T_d$ symmetry to $C_{3v}$.
In terms of these parameters the eigenvalues in  Eq.~\eqref{eq:Hcr} are
\begin{align}
  \epsilon_i - \epsilon_3 = \frac{\Delta_c}{2} \Bigg[ & \frac{(-1)^i}{3} \operatorname{sign}\left(1 - \frac{3 K}{\Delta_c}\right) \sqrt{8 + \left( 1 - \frac{3K}{\Delta_c} \right)^2} \notag \\
  \label{eq:HcrEnergies}
  & - 1  - \frac{K}{\Delta_{c}} + \frac{K'}{\Delta_{c}} \Bigg],
\end{align}
and the mixing angle, as used in Eq.~\eqref{eq:HcrStates}, is
\begin{align}
  \label{eq:mixangle}
  \phi = \frac{1}{2} \operatorname{arctan} \left( \frac{2 \sqrt{2}}{1 - 3K/\Delta_c} \right) .
\end{align}

\section{Basis change from \texorpdfstring{$T_d$}{Td} to \texorpdfstring{$C_{3v}$}{C3v} basis\label{sec:LTd}}

Considering that the nearest neighboring atoms of the TM defect respects $T_d$ symmetry, it may be justified to only take the leading order of the symmetry reduction into account, e.g., by assuming that only $H_{o}$ reduces the $T_d$ to $C_{3v}$ symmetry.
With this assumption one can use that intact $T_d$ symmetry has fewer independent parameters compared to $C_{3v}$. 
The reduction factors introduced in the main text are further restricted for intact $T_d$ symmetry, as they need to fulfill $p_{k,11} = 0$, $p_{k,i2} = p_{k,i3}$ and $\tilde{p}_{k,ij} = \tilde{p}_{k',ij}$ independently of the direction $k$ for $p = \lambda, r, \mathcal{E} $. For the crystal potential parameters, this implies $\phi_{T_d} = \operatorname{atan}(2 \sqrt{2})/2 = \operatorname{atan}({1}/{\sqrt{2}})$ and $\epsilon_2 = \epsilon_3$.

The basis states of a crystal potential that leaves $T_d$ symmetry intact and one that reduces it to $C_{3v}$ are related by
\begin{align}
\ket{\pm_1} = & \cos{\left(\phi_{T_d} - \phi \right)}\ket{\widetilde{\pm_1}} -  \sin{\left(\phi_{T_d} - \phi \right)}\ket{\widetilde{\pm_2}}, \\
\ket{\pm_2} = & \sin{\left(\phi_{T_d} - \phi \right)}\ket{\widetilde{\pm_1}} + \cos{\left(\phi_{T_d} - \phi_{\mathrm{cr}} \right)}\ket{\widetilde{\pm_2}}, \\
\ket{0} = & \ket{\widetilde{0}},
\end{align}
where we denote the eigenstates for the $T_d$ symmetric crystal potential by $\ket{\widetilde{\pm_i}}, \ket{\tilde{0}}$.
Using this transformation, orbital operators transforming according to $T_d$ symmetry can be easily transformed to the $C_{3v}$ basis using the mixing angle $\phi$.


\section{Parameters for V in the \texorpdfstring{$\alpha$}{α}-configuration of 
\texorpdfstring{$6H$}{6H}-SiC\label{sec:paramVSih6H}}
The necessary parameters for the model for vanadium (V) in the $\alpha$-configuration of $6H$-SiC were taken from fits to experimental data by Kaufmann \textit{et al.}~\cite{kaufmann97}.
While this has no implications for the employed theory, we mention that Kaufmann \textit{et al.} assumed that the $\alpha$-configuration of $6H$-SiC corresponds to a defect in a quasihexagonal layer, while recent \textit{ab initio} calculations~\cite{csore19} suggest that it corresponds to a defect in one of the quasicubic layers.
The parameters used in this article are
\begin{align}
  & \epsilon_3 = 1018.47 \, \mathrm{meV}, & 
    \Delta_c = 973.15 \, \mathrm{meV}, \\
  &   K=93.36 \, \mathrm{meV}, & 
    K'=-24.30 \, \mathrm{meV}.
\end{align}
Using Eq.~\eqref{eq:HcrEnergies}
and shifting the energy scale such that $\epsilon_1 = 0$
leads to the remaining parameters of Eqs.~\eqref{eq:Hcr}~and~\eqref{eq:HcrStates}, 
$\phi \approx 0.662$ and $\epsilon_2 \approx 946.14 \, \mathrm{meV}$.
The orbital operators in $H_{z}$ and $H_{\mathrm{so}}$ can be calculated as explained in Appendix~\ref{sec:LTd} from the $T_d$ symmetric factors
\begin{align}
  &  \tilde{r}_{12} = 0.18, & 
   \tilde{r}_{22} = 0.75, \\
  & \tilde{\lambda}_{12} = 1.77 \, \mathrm{meV}, & 
\tilde{\lambda}_{22} = 16.29 \, \mathrm{meV} .
\end{align}
Furthermore, the free ion value of the spin-orbit coupling strength for V is given by $\lambda_0 = 30.75 \, \mathrm{meV}$ in the same paper.
Finally, we use $g_s = 2$.

\section{Block Diagonal Basis and Analytic Spin-Orbit States\label{sec:AnSO}}
The Hamiltonian $H_{o} + H_{\mathrm{so}}$ is block diagonal in the following basis.
The first block is defined in the subspace spanned by the basis states
\begin{equation}
\frac{1}{\sqrt{2}}(\mathrm{i}\ket{+_1}\ket{\uparrow} + \ket{-_1}\ket{\downarrow}),
\frac{1}{\sqrt{2}}(\mathrm{i}\ket{+_2}\ket{\uparrow} + \ket{-_2}\ket{\downarrow}),
\end{equation}
that are related by time-inversion to the second blocks basis states
\begin{equation}
\frac{-1}{\sqrt{2}}(\mathrm{i}\ket{+_1}\ket{\uparrow} - \ket{-_1}\ket{\downarrow}),
\frac{-1}{\sqrt{2}}(\mathrm{i}\ket{+_2}\ket{\uparrow} - \ket{-_2}\ket{\downarrow}).
\end{equation}
Again, the third blocks basis states
\begin{equation}
\ket{+_1}\ket{\downarrow},
\ket{0}\ket{\uparrow},
\ket{+_2}\ket{\downarrow}
\end{equation}
are related to the fourth blocks basis states
\begin{equation}
- i\ket{-_1}\ket{\uparrow},
- i\ket{0}\ket{\downarrow},
- i\ket{-_2}\ket{\uparrow},
\end{equation}
by time inversion.
Because the maximal block size is $3\times3$ and the blocks of this size are real and symmetric it is possible to diagonalize the matrices analytically.
The eigenvalues are
\begin{widetext}
\begin{align}
  \label{eq:EVan}
  E_{i,\Gamma_{5/6}} = & \frac{1}{2} \left( \epsilon_1 + \frac{\lambda_{\parallel11}}{2} + \epsilon_2 + \frac{\lambda_{\parallel22}}{2} \right) + (-1)^i \frac{1}{2} \left(\epsilon_2 + \frac{\lambda_{\parallel22}}{2} - \epsilon_1 - \frac{\lambda_{\parallel11}}{2} \right) \sqrt{\frac{\lambda_{\parallel12}^{2} + 4 \lambda_{\perp12}^{2}}{\left(\epsilon_2 + \frac{\lambda_{\parallel22}}{2} - \epsilon_1 - \frac{\lambda_{\parallel11}}{2}\right)^{2}} + 1} , \\
  E_{i,\Gamma_{4}} = & \frac{b}{3} + 2 \sqrt{b^2 - 3c} \cos{\left[\frac{\Delta}{3} + \left( \delta_{i1} - \delta_{i2} \right) \frac{2\pi}{3} \right]} , \\
\text{with } &  \notag \left\{
  \begin{aligned}
  b = & \epsilon_0 + \epsilon_1 + \epsilon_2 - \frac{\lambda_{\parallel11}}{2} - \frac{\lambda_{\parallel22}}{2}, \qquad \cos(\Delta) = \frac{2 b^{3} - 9 b c - 27 d}{2 \sqrt{\left(b^{2} - 3 c\right)^{3}}}, \\
  c = & \epsilon_0 \left(\epsilon_1 + \epsilon_2 - \frac{\lambda_{\parallel11}}{2} - \frac{\lambda_{\parallel22}}{2}\right) - \frac{\lambda_{\parallel12}^{2}}{4} - \lambda_{\perp01}^{2} - \lambda_{\perp02}^{2} + \left(\epsilon_1 - \frac{\lambda_{\parallel11}}{2}\right) \left(\epsilon_2 - \frac{\lambda_{\parallel22}}{2}\right) , \\
  d = & \epsilon_0 \left[\frac{\lambda_{\parallel12}^{2}}{4} - \left(\epsilon_1 - \frac{\lambda_{\parallel11}}{2}\right) \left(\epsilon_2 - \frac{\lambda_{\parallel22}}{2}\right)\right] + \lambda_{\parallel12} \lambda_{\perp01} \lambda_{\perp02} + \lambda_{\perp01}^{2} \left(\epsilon_2 - \frac{\lambda_{\parallel22}}{2}\right) + \lambda_{\perp02}^{2} \left(\epsilon_1 - \frac{\lambda_{\parallel11}}{2}\right),
\end{aligned} \right.
\end{align}
\end{widetext}
where the labels are compatible with the perturbative solution.
Since the $3 \times 3$ blocks are real, the transformation can be expressed in terms of (three) Euler angles.
The diagonalization of the $2 \times 2$ blocks can expressed in terms of two angles.

\bibliography{refs.bib}

\begin{thebibliography}{32}%
\makeatletter
\providecommand \@ifxundefined [1]{%
 \@ifx{#1\undefined}
}%
\providecommand \@ifnum [1]{%
 \ifnum #1\expandafter \@firstoftwo
 \else \expandafter \@secondoftwo
 \fi
}%
\providecommand \@ifx [1]{%
 \ifx #1\expandafter \@firstoftwo
 \else \expandafter \@secondoftwo
 \fi
}%
\providecommand \natexlab [1]{#1}%
\providecommand \enquote  [1]{``#1''}%
\providecommand \bibnamefont  [1]{#1}%
\providecommand \bibfnamefont [1]{#1}%
\providecommand \citenamefont [1]{#1}%
\providecommand \href@noop [0]{\@secondoftwo}%
\providecommand \href [0]{\begingroup \@sanitize@url \@href}%
\providecommand \@href[1]{\@@startlink{#1}\@@href}%
\providecommand \@@href[1]{\endgroup#1\@@endlink}%
\providecommand \@sanitize@url [0]{\catcode `\\12\catcode `\$12\catcode
  `\&12\catcode `\#12\catcode `\^12\catcode `\_12\catcode `\%12\relax}%
\providecommand \@@startlink[1]{}%
\providecommand \@@endlink[0]{}%
\providecommand \url  [0]{\begingroup\@sanitize@url \@url }%
\providecommand \@url [1]{\endgroup\@href {#1}{\urlprefix }}%
\providecommand \urlprefix  [0]{URL }%
\providecommand \Eprint [0]{\href }%
\providecommand \doibase [0]{https://doi.org/}%
\providecommand \selectlanguage [0]{\@gobble}%
\providecommand \bibinfo  [0]{\@secondoftwo}%
\providecommand \bibfield  [0]{\@secondoftwo}%
\providecommand \translation [1]{[#1]}%
\providecommand \BibitemOpen [0]{}%
\providecommand \bibitemStop [0]{}%
\providecommand \bibitemNoStop [0]{.\EOS\space}%
\providecommand \EOS [0]{\spacefactor3000\relax}%
\providecommand \BibitemShut  [1]{\csname bibitem#1\endcsname}%
\let\auto@bib@innerbib\@empty
\bibitem [{\citenamefont {Gisin}\ and\ \citenamefont {Thew}(2007)}]{gisin07}%
  \BibitemOpen
  \bibfield  {author} {\bibinfo {author} {\bibfnamefont {N.}~\bibnamefont
  {Gisin}}\ and\ \bibinfo {author} {\bibfnamefont {R.}~\bibnamefont {Thew}},\
  }\bibfield  {title} {\bibinfo {title} {Quantum communication},\ }\href
  {https://doi.org/10.1038/nphoton.2007.22} {\bibfield  {journal} {\bibinfo
  {journal} {Nat. Photonics}\ }\textbf {\bibinfo {volume} {1}},\ \bibinfo
  {pages} {165} (\bibinfo {year} {2007})}\BibitemShut {NoStop}%
\bibitem [{\citenamefont {Kimble}(2008)}]{kimble08}%
  \BibitemOpen
  \bibfield  {author} {\bibinfo {author} {\bibfnamefont {H.~J.}\ \bibnamefont
  {Kimble}},\ }\bibfield  {title} {\bibinfo {title} {The quantum internet},\
  }\href {https://doi.org/10.1038/nature07127} {\bibfield  {journal} {\bibinfo
  {journal} {Nature}\ }\textbf {\bibinfo {volume} {453}},\ \bibinfo {pages}
  {1023} (\bibinfo {year} {2008})}\BibitemShut {NoStop}%
\bibitem [{\citenamefont {Spindlberger}\ \emph {et~al.}(2019)\citenamefont
  {Spindlberger}, \citenamefont {Cs\'or\'e}, \citenamefont {Thiering},
  \citenamefont {Putz}, \citenamefont {Karhu}, \citenamefont {Hassan},
  \citenamefont {Son}, \citenamefont {Fromherz}, \citenamefont {Gali},\ and\
  \citenamefont {Trupke}}]{spindlberger19}%
  \BibitemOpen
  \bibfield  {author} {\bibinfo {author} {\bibfnamefont {L.}~\bibnamefont
  {Spindlberger}}, \bibinfo {author} {\bibfnamefont {A.}~\bibnamefont
  {Cs\'or\'e}}, \bibinfo {author} {\bibfnamefont {G.}~\bibnamefont {Thiering}},
  \bibinfo {author} {\bibfnamefont {S.}~\bibnamefont {Putz}}, \bibinfo {author}
  {\bibfnamefont {R.}~\bibnamefont {Karhu}}, \bibinfo {author} {\bibfnamefont
  {J.}~\bibnamefont {Hassan}}, \bibinfo {author} {\bibfnamefont
  {N.}~\bibnamefont {Son}}, \bibinfo {author} {\bibfnamefont {T.}~\bibnamefont
  {Fromherz}}, \bibinfo {author} {\bibfnamefont {A.}~\bibnamefont {Gali}},\
  and\ \bibinfo {author} {\bibfnamefont {M.}~\bibnamefont {Trupke}},\
  }\bibfield  {title} {\bibinfo {title} {Optical properties of vanadium in 4{H}
  silicon carbide for quantum technology},\ }\href
  {https://doi.org/10.1103/physrevapplied.12.014015} {\bibfield  {journal}
  {\bibinfo  {journal} {Phys. Rev. Appl.}\ }\textbf {\bibinfo {volume} {12}},\
  \bibinfo {pages} {014015} (\bibinfo {year} {2019})}\BibitemShut {NoStop}%
\bibitem [{\citenamefont {Wolfowicz}\ \emph {et~al.}(2020)\citenamefont
  {Wolfowicz}, \citenamefont {Anderson}, \citenamefont {Diler}, \citenamefont
  {Poluektov}, \citenamefont {Heremans},\ and\ \citenamefont
  {Awschalom}}]{wolfowicz20}%
  \BibitemOpen
  \bibfield  {author} {\bibinfo {author} {\bibfnamefont {G.}~\bibnamefont
  {Wolfowicz}}, \bibinfo {author} {\bibfnamefont {C.~P.}\ \bibnamefont
  {Anderson}}, \bibinfo {author} {\bibfnamefont {B.}~\bibnamefont {Diler}},
  \bibinfo {author} {\bibfnamefont {O.~G.}\ \bibnamefont {Poluektov}}, \bibinfo
  {author} {\bibfnamefont {F.~J.}\ \bibnamefont {Heremans}},\ and\ \bibinfo
  {author} {\bibfnamefont {D.~D.}\ \bibnamefont {Awschalom}},\ }\bibfield
  {title} {\bibinfo {title} {Vanadium spin qubits as telecom quantum emitters
  in silicon carbide},\ }\href {https://doi.org/10.1126/sciadv.aaz1192}
  {\bibfield  {journal} {\bibinfo  {journal} {Sci. Adv.}\ }\textbf {\bibinfo
  {volume} {6}},\ \bibinfo {pages} {eaaz1192} (\bibinfo {year}
  {2020})}\BibitemShut {NoStop}%
\bibitem [{\citenamefont {Gilardoni}\ \emph {et~al.}(2020)\citenamefont
  {Gilardoni}, \citenamefont {Bosma}, \citenamefont {Hien}, \citenamefont
  {Hendriks}, \citenamefont {Magnusson}, \citenamefont {Ellison}, \citenamefont
  {Ivanov}, \citenamefont {Son},\ and\ \citenamefont {Wal}}]{gilardoni20}%
  \BibitemOpen
  \bibfield  {author} {\bibinfo {author} {\bibfnamefont {C.~M.}\ \bibnamefont
  {Gilardoni}}, \bibinfo {author} {\bibfnamefont {T.}~\bibnamefont {Bosma}},
  \bibinfo {author} {\bibfnamefont {D.~v.}\ \bibnamefont {Hien}}, \bibinfo
  {author} {\bibfnamefont {F.}~\bibnamefont {Hendriks}}, \bibinfo {author}
  {\bibfnamefont {B.}~\bibnamefont {Magnusson}}, \bibinfo {author}
  {\bibfnamefont {A.}~\bibnamefont {Ellison}}, \bibinfo {author} {\bibfnamefont
  {I.~G.}\ \bibnamefont {Ivanov}}, \bibinfo {author} {\bibfnamefont {N.~T.}\
  \bibnamefont {Son}},\ and\ \bibinfo {author} {\bibfnamefont {C.~H. v.~d.}\
  \bibnamefont {Wal}},\ }\bibfield  {title} {\bibinfo {title} {Spin-relaxation
  times exceeding seconds for color centers with strong spin–orbit coupling
  in {SiC}},\ }\href {https://doi.org/10.1088/1367-2630/abbf23} {\bibfield
  {journal} {\bibinfo  {journal} {New J. Phys.}\ }\textbf {\bibinfo {volume}
  {22}},\ \bibinfo {pages} {103051} (\bibinfo {year} {2020})}\BibitemShut
  {NoStop}%
\bibitem [{\citenamefont {Bosma}\ \emph {et~al.}(2018)\citenamefont {Bosma},
  \citenamefont {Lof}, \citenamefont {Gilardoni}, \citenamefont {Zwier},
  \citenamefont {Hendriks}, \citenamefont {Magnusson}, \citenamefont {Ellison},
  \citenamefont {G\"allstr\"om}, \citenamefont {Ivanov}, \citenamefont {Son},
  \citenamefont {Havenith},\ and\ \citenamefont {van~der Wal}}]{bosma18}%
  \BibitemOpen
  \bibfield  {author} {\bibinfo {author} {\bibfnamefont {T.}~\bibnamefont
  {Bosma}}, \bibinfo {author} {\bibfnamefont {G.~J.~J.}\ \bibnamefont {Lof}},
  \bibinfo {author} {\bibfnamefont {C.~M.}\ \bibnamefont {Gilardoni}}, \bibinfo
  {author} {\bibfnamefont {O.~V.}\ \bibnamefont {Zwier}}, \bibinfo {author}
  {\bibfnamefont {F.}~\bibnamefont {Hendriks}}, \bibinfo {author}
  {\bibfnamefont {B.}~\bibnamefont {Magnusson}}, \bibinfo {author}
  {\bibfnamefont {A.}~\bibnamefont {Ellison}}, \bibinfo {author} {\bibfnamefont
  {A.}~\bibnamefont {G\"allstr\"om}}, \bibinfo {author} {\bibfnamefont {I.~G.}\
  \bibnamefont {Ivanov}}, \bibinfo {author} {\bibfnamefont {N.~T.}\
  \bibnamefont {Son}}, \bibinfo {author} {\bibfnamefont {R.~W.~A.}\
  \bibnamefont {Havenith}},\ and\ \bibinfo {author} {\bibfnamefont {C.~H.}\
  \bibnamefont {van~der Wal}},\ }\bibfield  {title} {\bibinfo {title}
  {Identification and tunable optical coherent control of transition-metal
  spins in silicon carbide},\ }\href
  {https://doi.org/10.1038/s41534-018-0097-8} {\bibfield  {journal} {\bibinfo
  {journal} {npj Quantum Inf.}\ }\textbf {\bibinfo {volume} {4}},\ \bibinfo
  {pages} {48} (\bibinfo {year} {2018})}\BibitemShut {NoStop}%
\bibitem [{\citenamefont {Kunzer}\ \emph {et~al.}(1993)\citenamefont {Kunzer},
  \citenamefont {M\"uller},\ and\ \citenamefont {Kaufmann}}]{kunzer93}%
  \BibitemOpen
  \bibfield  {author} {\bibinfo {author} {\bibfnamefont {M.}~\bibnamefont
  {Kunzer}}, \bibinfo {author} {\bibfnamefont {H.~D.}\ \bibnamefont
  {M\"uller}},\ and\ \bibinfo {author} {\bibfnamefont {U.}~\bibnamefont
  {Kaufmann}},\ }\bibfield  {title} {\bibinfo {title} {Magnetic circular
  dichroism and site-selective optically detected magnetic resonance of the
  deep amphoteric vanadium impurity in {6H}-{SiC}},\ }\href
  {https://doi.org/10.1103/physrevb.48.10846} {\bibfield  {journal} {\bibinfo
  {journal} {Phys. Rev. B}\ }\textbf {\bibinfo {volume} {48}},\ \bibinfo
  {pages} {10846} (\bibinfo {year} {1993})}\BibitemShut {NoStop}%
\bibitem [{\citenamefont {Reinke}\ \emph {et~al.}(1993)\citenamefont {Reinke},
  \citenamefont {Weihrich}, \citenamefont {Greulich-Weber},\ and\ \citenamefont
  {Spaeth}}]{reinke93}%
  \BibitemOpen
  \bibfield  {author} {\bibinfo {author} {\bibfnamefont {J.}~\bibnamefont
  {Reinke}}, \bibinfo {author} {\bibfnamefont {H.}~\bibnamefont {Weihrich}},
  \bibinfo {author} {\bibfnamefont {S.}~\bibnamefont {Greulich-Weber}},\ and\
  \bibinfo {author} {\bibfnamefont {J.-M.}\ \bibnamefont {Spaeth}},\ }\bibfield
   {title} {\bibinfo {title} {Magnetic circular dichroism of a vanadium
  impurity in {6H}-silicon carbide},\ }\href
  {https://doi.org/10.1088/0268-1242/8/10/013} {\bibfield  {journal} {\bibinfo
  {journal} {Semicond. Sci. Technol.}\ }\textbf {\bibinfo {volume} {8}},\
  \bibinfo {pages} {1862} (\bibinfo {year} {1993})}\BibitemShut {NoStop}%
\bibitem [{\citenamefont {Kaufmann}\ \emph {et~al.}(1997)\citenamefont
  {Kaufmann}, \citenamefont {D\"ornen},\ and\ \citenamefont
  {Ham}}]{kaufmann97}%
  \BibitemOpen
  \bibfield  {author} {\bibinfo {author} {\bibfnamefont {B.}~\bibnamefont
  {Kaufmann}}, \bibinfo {author} {\bibfnamefont {A.}~\bibnamefont {D\"ornen}},\
  and\ \bibinfo {author} {\bibfnamefont {F.~S.}\ \bibnamefont {Ham}},\
  }\bibfield  {title} {\bibinfo {title} {Crystal-field model of vanadium in
  {6H} silicon carbide},\ }\href {https://doi.org/10.1103/physrevb.55.13009}
  {\bibfield  {journal} {\bibinfo  {journal} {Phys. Rev. B}\ }\textbf {\bibinfo
  {volume} {55}},\ \bibinfo {pages} {13009} (\bibinfo {year}
  {1997})}\BibitemShut {NoStop}%
\bibitem [{\citenamefont {Cs\'or\'e}\ and\ \citenamefont
  {Gali}(2019)}]{csore19}%
  \BibitemOpen
  \bibfield  {author} {\bibinfo {author} {\bibfnamefont {A.}~\bibnamefont
  {Cs\'or\'e}}\ and\ \bibinfo {author} {\bibfnamefont {A.}~\bibnamefont
  {Gali}},\ }\bibfield  {title} {\bibinfo {title} {Ab initio determination of
  pseudospin for paramagnetic defects in {SiC}},\ }\Eprint
  {https://arxiv.org/abs/1909.11587} {arXiv:1909.11587 [quant-ph]}  (\bibinfo
  {year} {2019})\BibitemShut {NoStop}%
\bibitem [{\citenamefont {Awschalom}\ \emph {et~al.}(2018)\citenamefont
  {Awschalom}, \citenamefont {Hanson}, \citenamefont {Wrachtrup},\ and\
  \citenamefont {Zhou}}]{awschalom18}%
  \BibitemOpen
  \bibfield  {author} {\bibinfo {author} {\bibfnamefont {D.~D.}\ \bibnamefont
  {Awschalom}}, \bibinfo {author} {\bibfnamefont {R.}~\bibnamefont {Hanson}},
  \bibinfo {author} {\bibfnamefont {J.}~\bibnamefont {Wrachtrup}},\ and\
  \bibinfo {author} {\bibfnamefont {B.~B.}\ \bibnamefont {Zhou}},\ }\bibfield
  {title} {\bibinfo {title} {Quantum technologies with optically interfaced
  solid-state spins},\ }\href {https://doi.org/10.1038/s41566-018-0232-2}
  {\bibfield  {journal} {\bibinfo  {journal} {Nat. Photonics}\ }\textbf
  {\bibinfo {volume} {12}},\ \bibinfo {pages} {516} (\bibinfo {year}
  {2018})}\BibitemShut {NoStop}%
\bibitem [{\citenamefont {Lenef}\ and\ \citenamefont {Rand}(1996)}]{lenef96}%
  \BibitemOpen
  \bibfield  {author} {\bibinfo {author} {\bibfnamefont {A.}~\bibnamefont
  {Lenef}}\ and\ \bibinfo {author} {\bibfnamefont {S.~C.}\ \bibnamefont
  {Rand}},\ }\bibfield  {title} {\bibinfo {title} {Electronic structure of the
  {N-V} center in diamond: Theory},\ }\href
  {https://doi.org/10.1103/physrevb.53.13441} {\bibfield  {journal} {\bibinfo
  {journal} {Phys. Rev. B}\ }\textbf {\bibinfo {volume} {53}},\ \bibinfo
  {pages} {13441} (\bibinfo {year} {1996})}\BibitemShut {NoStop}%
\bibitem [{\citenamefont {Tamarat}\ \emph {et~al.}(2008)\citenamefont
  {Tamarat}, \citenamefont {Manson}, \citenamefont {Harrison}, \citenamefont
  {McMurtrie}, \citenamefont {Nizovtsev}, \citenamefont {Santori},
  \citenamefont {Beausoleil}, \citenamefont {Neumann}, \citenamefont {Gaebel},
  \citenamefont {Jelezko}, \citenamefont {Hemmer},\ and\ \citenamefont
  {Wrachtrup}}]{tamarat08}%
  \BibitemOpen
  \bibfield  {author} {\bibinfo {author} {\bibfnamefont {P.}~\bibnamefont
  {Tamarat}}, \bibinfo {author} {\bibfnamefont {N.~B.}\ \bibnamefont {Manson}},
  \bibinfo {author} {\bibfnamefont {J.~P.}\ \bibnamefont {Harrison}}, \bibinfo
  {author} {\bibfnamefont {R.~L.}\ \bibnamefont {McMurtrie}}, \bibinfo {author}
  {\bibfnamefont {A.}~\bibnamefont {Nizovtsev}}, \bibinfo {author}
  {\bibfnamefont {C.}~\bibnamefont {Santori}}, \bibinfo {author} {\bibfnamefont
  {R.~G.}\ \bibnamefont {Beausoleil}}, \bibinfo {author} {\bibfnamefont
  {P.}~\bibnamefont {Neumann}}, \bibinfo {author} {\bibfnamefont
  {T.}~\bibnamefont {Gaebel}}, \bibinfo {author} {\bibfnamefont
  {F.}~\bibnamefont {Jelezko}}, \bibinfo {author} {\bibfnamefont
  {P.}~\bibnamefont {Hemmer}},\ and\ \bibinfo {author} {\bibfnamefont
  {J.}~\bibnamefont {Wrachtrup}},\ }\bibfield  {title} {\bibinfo {title}
  {Spin-flip and spin-conserving optical transitions of the nitrogen-vacancy
  centre in diamond},\ }\href {https://doi.org/10.1088/1367-2630/10/4/045004}
  {\bibfield  {journal} {\bibinfo  {journal} {New J. Phys.}\ }\textbf {\bibinfo
  {volume} {10}},\ \bibinfo {pages} {045004} (\bibinfo {year}
  {2008})}\BibitemShut {NoStop}%
\bibitem [{\citenamefont {Doherty}\ \emph {et~al.}(2011)\citenamefont
  {Doherty}, \citenamefont {Manson}, \citenamefont {Delaney},\ and\
  \citenamefont {Hollenberg}}]{doherty11}%
  \BibitemOpen
  \bibfield  {author} {\bibinfo {author} {\bibfnamefont {M.~W.}\ \bibnamefont
  {Doherty}}, \bibinfo {author} {\bibfnamefont {N.~B.}\ \bibnamefont {Manson}},
  \bibinfo {author} {\bibfnamefont {P.}~\bibnamefont {Delaney}},\ and\ \bibinfo
  {author} {\bibfnamefont {L.~C.~L.}\ \bibnamefont {Hollenberg}},\ }\bibfield
  {title} {\bibinfo {title} {The negatively charged nitrogen-vacancy centre in
  diamond: the electronic solution},\ }\href
  {https://doi.org/10.1088/1367-2630/13/2/025019} {\bibfield  {journal}
  {\bibinfo  {journal} {New J. Phys.}\ }\textbf {\bibinfo {volume} {13}},\
  \bibinfo {pages} {025019} (\bibinfo {year} {2011})}\BibitemShut {NoStop}%
\bibitem [{\citenamefont {Maze}\ \emph {et~al.}(2011)\citenamefont {Maze},
  \citenamefont {Gali}, \citenamefont {Togan}, \citenamefont {Chu},
  \citenamefont {Trifonov}, \citenamefont {Kaxiras},\ and\ \citenamefont
  {Lukin}}]{maze11}%
  \BibitemOpen
  \bibfield  {author} {\bibinfo {author} {\bibfnamefont {J.~R.}\ \bibnamefont
  {Maze}}, \bibinfo {author} {\bibfnamefont {A.}~\bibnamefont {Gali}}, \bibinfo
  {author} {\bibfnamefont {E.}~\bibnamefont {Togan}}, \bibinfo {author}
  {\bibfnamefont {Y.}~\bibnamefont {Chu}}, \bibinfo {author} {\bibfnamefont
  {A.}~\bibnamefont {Trifonov}}, \bibinfo {author} {\bibfnamefont
  {E.}~\bibnamefont {Kaxiras}},\ and\ \bibinfo {author} {\bibfnamefont {M.~D.}\
  \bibnamefont {Lukin}},\ }\bibfield  {title} {\bibinfo {title} {Properties of
  nitrogen-vacancy centers in diamond: the group theoretic approach},\ }\href
  {https://doi.org/10.1088/1367-2630/13/2/025025} {\bibfield  {journal}
  {\bibinfo  {journal} {New J. Phys.}\ }\textbf {\bibinfo {volume} {13}},\
  \bibinfo {pages} {025025} (\bibinfo {year} {2011})}\BibitemShut {NoStop}%
\bibitem [{\citenamefont {Doherty}\ \emph {et~al.}(2013)\citenamefont
  {Doherty}, \citenamefont {Manson}, \citenamefont {Delaney}, \citenamefont
  {Jelezko}, \citenamefont {Wrachtrup},\ and\ \citenamefont
  {Hollenberg}}]{doherty13}%
  \BibitemOpen
  \bibfield  {author} {\bibinfo {author} {\bibfnamefont {M.~W.}\ \bibnamefont
  {Doherty}}, \bibinfo {author} {\bibfnamefont {N.~B.}\ \bibnamefont {Manson}},
  \bibinfo {author} {\bibfnamefont {P.}~\bibnamefont {Delaney}}, \bibinfo
  {author} {\bibfnamefont {F.}~\bibnamefont {Jelezko}}, \bibinfo {author}
  {\bibfnamefont {J.}~\bibnamefont {Wrachtrup}},\ and\ \bibinfo {author}
  {\bibfnamefont {L.~C.}\ \bibnamefont {Hollenberg}},\ }\bibfield  {title}
  {\bibinfo {title} {The nitrogen-vacancy colour centre in diamond},\ }\href
  {https://doi.org/10.1016/j.physrep.2013.02.001} {\bibfield  {journal}
  {\bibinfo  {journal} {Phys. Rep.}\ }\textbf {\bibinfo {volume} {528}},\
  \bibinfo {pages} {1} (\bibinfo {year} {2013})}\BibitemShut {NoStop}%
\bibitem [{\citenamefont {Dietz}\ \emph {et~al.}(1963)\citenamefont {Dietz},
  \citenamefont {Kamimura}, \citenamefont {Sturge},\ and\ \citenamefont
  {Yariv}}]{dietz63}%
  \BibitemOpen
  \bibfield  {author} {\bibinfo {author} {\bibfnamefont {R.~E.}\ \bibnamefont
  {Dietz}}, \bibinfo {author} {\bibfnamefont {H.}~\bibnamefont {Kamimura}},
  \bibinfo {author} {\bibfnamefont {M.~D.}\ \bibnamefont {Sturge}},\ and\
  \bibinfo {author} {\bibfnamefont {A.}~\bibnamefont {Yariv}},\ }\bibfield
  {title} {\bibinfo {title} {Electronic structure of copper impurities in
  {ZnO}},\ }\href {https://doi.org/10.1103/physrev.132.1559} {\bibfield
  {journal} {\bibinfo  {journal} {Phys. Rev.}\ }\textbf {\bibinfo {volume}
  {132}},\ \bibinfo {pages} {1559} (\bibinfo {year} {1963})}\BibitemShut
  {NoStop}%
\bibitem [{\citenamefont {Dresselhaus}\ \emph {et~al.}(2010)\citenamefont
  {Dresselhaus}, \citenamefont {Dresselhaus},\ and\ \citenamefont
  {Jorio}}]{dresselhaus10}%
  \BibitemOpen
  \bibfield  {author} {\bibinfo {author} {\bibfnamefont {M.~S.}\ \bibnamefont
  {Dresselhaus}}, \bibinfo {author} {\bibfnamefont {G.}~\bibnamefont
  {Dresselhaus}},\ and\ \bibinfo {author} {\bibfnamefont {A.}~\bibnamefont
  {Jorio}},\ }\href@noop {} {\emph {\bibinfo {title} {Group theory :
  application to the physics of condensed matter}}}\ (\bibinfo  {publisher}
  {Springer-Verlag},\ \bibinfo {address} {Berlin},\ \bibinfo {year}
  {2010})\BibitemShut {NoStop}%
\bibitem [{\citenamefont {Thomas}(1926)}]{thomas26}%
  \BibitemOpen
  \bibfield  {author} {\bibinfo {author} {\bibfnamefont {L.~H.}\ \bibnamefont
  {Thomas}},\ }\bibfield  {title} {\bibinfo {title} {The motion of the spinning
  electron},\ }\href {https://doi.org/10.1038/117514a0} {\bibfield  {journal}
  {\bibinfo  {journal} {Nature}\ }\textbf {\bibinfo {volume} {117}},\ \bibinfo
  {pages} {514} (\bibinfo {year} {1926})}\BibitemShut {NoStop}%
\bibitem [{\citenamefont {Coish}\ and\ \citenamefont {Baugh}(2009)}]{coish09}%
  \BibitemOpen
  \bibfield  {author} {\bibinfo {author} {\bibfnamefont {W.~A.}\ \bibnamefont
  {Coish}}\ and\ \bibinfo {author} {\bibfnamefont {J.}~\bibnamefont {Baugh}},\
  }\bibfield  {title} {\bibinfo {title} {Nuclear spins in nanostructures},\
  }\href {https://doi.org/10.1002/pssb.200945229} {\bibfield  {journal}
  {\bibinfo  {journal} {physica status solidi (b)}\ }\textbf {\bibinfo {volume}
  {246}},\ \bibinfo {pages} {2203} (\bibinfo {year} {2009})}\BibitemShut
  {NoStop}%
\bibitem [{\citenamefont {Audi}\ \emph {et~al.}(2003)\citenamefont {Audi},
  \citenamefont {Bersillon}, \citenamefont {Blachot},\ and\ \citenamefont
  {Wapstra}}]{audi03}%
  \BibitemOpen
  \bibfield  {author} {\bibinfo {author} {\bibfnamefont {G.}~\bibnamefont
  {Audi}}, \bibinfo {author} {\bibfnamefont {O.}~\bibnamefont {Bersillon}},
  \bibinfo {author} {\bibfnamefont {J.}~\bibnamefont {Blachot}},\ and\ \bibinfo
  {author} {\bibfnamefont {A.}~\bibnamefont {Wapstra}},\ }\bibfield  {title}
  {\bibinfo {title} {The {NUBASE} evaluation of nuclear and decay properties},\
  }\href {https://doi.org/10.1016/j.nuclphysa.2003.11.001} {\bibfield
  {journal} {\bibinfo  {journal} {Nucl. Phys. A}\ }\textbf {\bibinfo {volume}
  {729}},\ \bibinfo {pages} {3} (\bibinfo {year} {2003})}\BibitemShut {NoStop}%
\bibitem [{\citenamefont {Meija}\ \emph {et~al.}(2016)\citenamefont {Meija},
  \citenamefont {Coplen}, \citenamefont {Berglund}, \citenamefont {Brand},
  \citenamefont {Bièvre}, \citenamefont {Gr\"{o}ning}, \citenamefont {Holden},
  \citenamefont {Irrgeher}, \citenamefont {Loss}, \citenamefont {Walczyk},\
  and\ \citenamefont {Prohaska}}]{meija16}%
  \BibitemOpen
  \bibfield  {author} {\bibinfo {author} {\bibfnamefont {J.}~\bibnamefont
  {Meija}}, \bibinfo {author} {\bibfnamefont {T.~B.}\ \bibnamefont {Coplen}},
  \bibinfo {author} {\bibfnamefont {M.}~\bibnamefont {Berglund}}, \bibinfo
  {author} {\bibfnamefont {W.~A.}\ \bibnamefont {Brand}}, \bibinfo {author}
  {\bibfnamefont {P.~D.}\ \bibnamefont {Bièvre}}, \bibinfo {author}
  {\bibfnamefont {M.}~\bibnamefont {Gr\"{o}ning}}, \bibinfo {author}
  {\bibfnamefont {N.~E.}\ \bibnamefont {Holden}}, \bibinfo {author}
  {\bibfnamefont {J.}~\bibnamefont {Irrgeher}}, \bibinfo {author}
  {\bibfnamefont {R.~D.}\ \bibnamefont {Loss}}, \bibinfo {author}
  {\bibfnamefont {T.}~\bibnamefont {Walczyk}},\ and\ \bibinfo {author}
  {\bibfnamefont {T.}~\bibnamefont {Prohaska}},\ }\bibfield  {title} {\bibinfo
  {title} {Isotopic compositions of the elements 2013 ({IUPAC} technical
  report)},\ }\href {https://doi.org/10.1515/pac-2015-0503} {\bibfield
  {journal} {\bibinfo  {journal} {Pure Appl. Chem.}\ }\textbf {\bibinfo
  {volume} {88}},\ \bibinfo {pages} {293} (\bibinfo {year} {2016})}\BibitemShut
  {NoStop}%
\bibitem [{\citenamefont {Koseki}\ \emph {et~al.}(2019)\citenamefont {Koseki},
  \citenamefont {Matsunaga}, \citenamefont {Asada}, \citenamefont {Schmidt},\
  and\ \citenamefont {Gordon}}]{koseki19}%
  \BibitemOpen
  \bibfield  {author} {\bibinfo {author} {\bibfnamefont {S.}~\bibnamefont
  {Koseki}}, \bibinfo {author} {\bibfnamefont {N.}~\bibnamefont {Matsunaga}},
  \bibinfo {author} {\bibfnamefont {T.}~\bibnamefont {Asada}}, \bibinfo
  {author} {\bibfnamefont {M.~W.}\ \bibnamefont {Schmidt}},\ and\ \bibinfo
  {author} {\bibfnamefont {M.~S.}\ \bibnamefont {Gordon}},\ }\bibfield  {title}
  {\bibinfo {title} {Spin-orbit coupling constants in atoms and ions of
  transition elements: Comparison of effective core potentials, model core
  potentials, and all-electron methods},\ }\href
  {https://doi.org/10.1021/acs.jpca.8b09218} {\bibfield  {journal} {\bibinfo
  {journal} {J. Phys. Chem. A}\ }\textbf {\bibinfo {volume} {123}},\ \bibinfo
  {pages} {2325} (\bibinfo {year} {2019})}\BibitemShut {NoStop}%
\bibitem [{\citenamefont {Stone}(2005)}]{stone05}%
  \BibitemOpen
  \bibfield  {author} {\bibinfo {author} {\bibfnamefont {N.}~\bibnamefont
  {Stone}},\ }\bibfield  {title} {\bibinfo {title} {Table of nuclear magnetic
  dipole and electric quadrupole moments},\ }\href
  {https://doi.org/10.1016/j.adt.2005.04.001} {\bibfield  {journal} {\bibinfo
  {journal} {At. Data Nucl. Data Tables}\ }\textbf {\bibinfo {volume} {90}},\
  \bibinfo {pages} {75} (\bibinfo {year} {2005})}\BibitemShut {NoStop}%
\bibitem [{\citenamefont {Yamaguchi}\ \emph {et~al.}(2018)\citenamefont
  {Yamaguchi}, \citenamefont {Kobayashi}, \citenamefont {Yamamoto},\ and\
  \citenamefont {Hirose}}]{yamaguchi18}%
  \BibitemOpen
  \bibfield  {author} {\bibinfo {author} {\bibfnamefont {K.}~\bibnamefont
  {Yamaguchi}}, \bibinfo {author} {\bibfnamefont {D.}~\bibnamefont
  {Kobayashi}}, \bibinfo {author} {\bibfnamefont {T.}~\bibnamefont
  {Yamamoto}},\ and\ \bibinfo {author} {\bibfnamefont {K.}~\bibnamefont
  {Hirose}},\ }\bibfield  {title} {\bibinfo {title} {Theoretical investigation
  of the breakdown electric field of {SiC} polymorphs},\ }\href
  {https://doi.org/10.1016/j.physb.2017.03.042} {\bibfield  {journal} {\bibinfo
   {journal} {Physica B}\ }\textbf {\bibinfo {volume} {532}},\ \bibinfo {pages}
  {99} (\bibinfo {year} {2018})}\BibitemShut {NoStop}%
\bibitem [{\citenamefont {Morko\c{c}}\ \emph {et~al.}(1994)\citenamefont
  {Morko\c{c}}, \citenamefont {Strite}, \citenamefont {Gao}, \citenamefont
  {Lin}, \citenamefont {Sverdlov},\ and\ \citenamefont {Burns}}]{morkoc94}%
  \BibitemOpen
  \bibfield  {author} {\bibinfo {author} {\bibfnamefont {H.}~\bibnamefont
  {Morko\c{c}}}, \bibinfo {author} {\bibfnamefont {S.}~\bibnamefont {Strite}},
  \bibinfo {author} {\bibfnamefont {G.~B.}\ \bibnamefont {Gao}}, \bibinfo
  {author} {\bibfnamefont {M.~E.}\ \bibnamefont {Lin}}, \bibinfo {author}
  {\bibfnamefont {B.}~\bibnamefont {Sverdlov}},\ and\ \bibinfo {author}
  {\bibfnamefont {M.}~\bibnamefont {Burns}},\ }\bibfield  {title} {\bibinfo
  {title} {Large‐band‐gap {SiC}, {III‐V} nitride, and {II‐VI}
  {ZnSe}‐based semiconductor device technologies},\ }\href
  {https://doi.org/10.1063/1.358463} {\bibfield  {journal} {\bibinfo  {journal}
  {J. Appl. Phys.}\ }\textbf {\bibinfo {volume} {76}},\ \bibinfo {pages} {1363}
  (\bibinfo {year} {1994})}\BibitemShut {NoStop}%
\bibitem [{\citenamefont {Ham}(1965)}]{ham65}%
  \BibitemOpen
  \bibfield  {author} {\bibinfo {author} {\bibfnamefont {F.~S.}\ \bibnamefont
  {Ham}},\ }\bibfield  {title} {\bibinfo {title} {Dynamical {J}ahn-{T}eller
  effect in paramagnetic resonance spectra: Orbital reduction factors and
  partial quenching of spin-orbit interaction},\ }\href
  {https://doi.org/10.1103/physrev.138.a1727} {\bibfield  {journal} {\bibinfo
  {journal} {Phys. Rev.}\ }\textbf {\bibinfo {volume} {138}},\ \bibinfo {pages}
  {A1727} (\bibinfo {year} {1965})}\BibitemShut {NoStop}%
\bibitem [{\citenamefont {Ham}(1968)}]{ham68}%
  \BibitemOpen
  \bibfield  {author} {\bibinfo {author} {\bibfnamefont {F.~S.}\ \bibnamefont
  {Ham}},\ }\bibfield  {title} {\bibinfo {title} {Effect of linear
  {J}ahn-{T}eller coupling on paramagnetic resonance in a $^2{E}$ state},\
  }\href {https://doi.org/10.1103/physrev.166.307} {\bibfield  {journal}
  {\bibinfo  {journal} {Phys. Rev.}\ }\textbf {\bibinfo {volume} {166}},\
  \bibinfo {pages} {307} (\bibinfo {year} {1968})}\BibitemShut {NoStop}%
\bibitem [{\citenamefont {Cornwell}(1997)}]{cornwell97}%
  \BibitemOpen
  \bibfield  {author} {\bibinfo {author} {\bibfnamefont {J.~F.}\ \bibnamefont
  {Cornwell}},\ }\href@noop {} {\emph {\bibinfo {title} {Group theory in
  physics : an introduction}}}\ (\bibinfo  {publisher} {Academic Press},\
  \bibinfo {address} {San Diego, Calif},\ \bibinfo {year} {1997})\BibitemShut
  {NoStop}%
\bibitem [{\citenamefont {Bravyi}\ \emph {et~al.}(2011)\citenamefont {Bravyi},
  \citenamefont {DiVincenzo},\ and\ \citenamefont {Loss}}]{bravyi11}%
  \BibitemOpen
  \bibfield  {author} {\bibinfo {author} {\bibfnamefont {S.}~\bibnamefont
  {Bravyi}}, \bibinfo {author} {\bibfnamefont {D.~P.}\ \bibnamefont
  {DiVincenzo}},\ and\ \bibinfo {author} {\bibfnamefont {D.}~\bibnamefont
  {Loss}},\ }\bibfield  {title} {\bibinfo {title} {{Schrieffer-Wolff}
  transformation for quantum many-body systems},\ }\href
  {https://doi.org/10.1016/j.aop.2011.06.004} {\bibfield  {journal} {\bibinfo
  {journal} {Ann. Phys.}\ }\textbf {\bibinfo {volume} {326}},\ \bibinfo {pages}
  {2793} (\bibinfo {year} {2011})}\BibitemShut {NoStop}%
\bibitem [{\citenamefont {Tinkham}(2003)}]{tinkham03}%
  \BibitemOpen
  \bibfield  {author} {\bibinfo {author} {\bibfnamefont {M.}~\bibnamefont
  {Tinkham}},\ }\href@noop {} {\emph {\bibinfo {title} {Group theory and
  quantum mechanics}}}\ (\bibinfo  {publisher} {Dover Publications},\ \bibinfo
  {address} {Mineola, N.Y},\ \bibinfo {year} {2003})\BibitemShut {NoStop}%
\bibitem [{\citenamefont {Koster}(1963)}]{koster63}%
  \BibitemOpen
  \bibfield  {author} {\bibinfo {author} {\bibfnamefont {G.}~\bibnamefont
  {Koster}},\ }\href@noop {} {\emph {\bibinfo {title} {Properties of the
  thirty-two point groups}}}\ (\bibinfo  {publisher} {M.I.T. Press},\ \bibinfo
  {year} {1963})\BibitemShut {NoStop}%
\end{thebibliography}%

\end{document}